\documentclass[]{jfm}
\usepackage{graphicx}
\usepackage{newtxtext}
\usepackage{newtxmath}
\usepackage{natbib}
\usepackage{hyperref}
\usepackage{dcolumn}
\usepackage{bm}
\usepackage{gensymb}
\usepackage{tabularx}
\usepackage{color}
\usepackage{braket}
\hypersetup{colorlinks=true,urlcolor=blue,citecolor=blue}

\title{Natural convection in a vertical channel. Part 3. Bifurcations of many (additional) unstable periodic orbits and their dynamical relevance}

\shorttitle{Natural convection in a vertical channel. Part 3}
\shortauthor{Z. Zheng, L.S. Tuckerman and T.M. Schneider}

\author{Zheng Zheng\aff{1}\corresp{\email{zheng.zheng@epfl.ch}},                             Laurette S. Tuckerman\aff{2} 
        \and Tobias M. Schneider\aff{1}}

\affiliation{\aff{1} Emergent Complexity in Physical Systems Laboratory (ECPS), \'Ecole Polytechnique F\'ed\'erale de Lausanne, CH 1015 Lausanne, Switzerland
\aff{2} Physique et M\'ecanique des Milieux H\'et\'erog\`enes (PMMH), CNRS, ESPCI Paris, PSL University, Sorbonne Universit\'e, Universit\'e de Paris, 75005 Paris, France}

\begin{document}
\maketitle

\begin{abstract}
Vertical thermal convection system exhibits weak turbulence and spatio-temporally chaotic behaviour. In this system, we report seven equilibria and 26 periodic orbits, all new and linearly unstable. These orbits, together with four previously studied in \citet{Zheng2024part2} bring the number of periodic orbit branches computed so far to 30, all solutions to the fully non-linear three-dimensional Navier--Stokes equations. These new invariant solutions capture intricate spatio-temporal flow patterns including straight, oblique, wavy, skewed and distorted convection rolls, as well as bursts and defects in rolls. These interesting and important fluid mechanical processes in a small flow unit are shown to appear locally and instantaneously in a chaotic simulation in a large domain. Most of the solution branches show rich spatial and/or spatio-temporal symmetries. The bifurcation-theoretic organisation of these solutions is discussed; the bifurcation scenarios include Hopf, pitchfork, saddle--node, period-doubling, period-halving, global homoclinic and heteroclinic bifurcations, as well as isolas. These orbits are shown to be able to reconstruct statistically the core part of the attractor, and these results may pave the way to quantitatively describing transitional fluid turbulence using periodic orbit theory.
\end{abstract}


\section{Introduction}
\par The vertical thermal convection system is a classical and fundamental model in fluid dynamics, describing the motion of a fluid bounded by two vertical walls maintained at different temperatures. The fluid layer is driven by both buoyancy forces arising from horizontal temperature gradients and shear forces. Convection plays a crucial role in a wide range of natural and industrial processes, including atmospheric and oceanic circulation, mantle convection, and thermal management in engineering applications \citep{Kaushika2003, Arici2015}. A better fundamental understanding of thermal convection is thus essential for tackling a diverse array of academic and industrial challenges \citep{Bodenschatz2000, Ahlers2009heat, Lohse2024ultimate}.

\par Like the well-known and well-studied Rayleigh--B\'enard convection, vertical convection is also an ideal system for studying pattern formation above the onset of convection. Recent advances in experimental, numerical, and theoretical studies continue to refine our understanding of the stability, transition, turbulence, and heat transport mechanisms which are governed by the deterministic flow equations; these make vertical convection a subject of ongoing interest in fluid mechanics research. We refer readers to the introductions of the two previous papers in this series \citep{Zheng2024part1, Zheng2024part2} for a more complete literature review of this field.

\par Using the methodology described in \citet{Zheng2024part1, Zheng2024part2}, we consider the vertical convection system as a (very) high-dimensional non-linear dynamical system and employ the dynamical-systems-based approach to investigate the flow dynamics. This approach has been established as a paradigm to study transition to turbulence in various shear-dominated flows; see reviews in \citet{Kawahara2012, Graham2021} and references therein. Note that in our work, we sometimes use the word turbulence to refer to spatio-temporally chaotic dynamics instead of to a fully developed turbulent flow with an energy cascade across multiple spatial scales. Building on ideas by Smale, Ruelle, Bowen and Sinai, turbulence is sometimes viewed from a deterministic dynamical systems perspective in the infinite-dimensional phase space of the Navier--Stokes equations as a chaotic walk through a forest of non-chaotic invariant solutions (particularly equilibria and periodic orbits) \citep{Lanford1982}. While equilibria may reproduce characteristic features of the flow, they are time-independent and so the information contained within such solutions is limited. However, unstable periodic orbits are much more dynamically important and are believed to be transiently visited by a weakly turbulent flow, and to form the skeleton and building blocks of the chaotic dynamics of transitional turbulence \citep{cvitanovic1991periodic, Kawahara2001}.

\par In our vertical convection system, the control parameters are Rayleigh number ($Ra$) which is proportional to the temperature difference between the two walls and Prandtl number ($Pr$), which is the ratio between kinematic viscosity and thermal diffusivity. In addition to individual invariant solutions that are identified at fixed control parameters of the system, a bifurcation analysis via parametric continuations in one of the control parameters ($Ra$ in this work) may reveal the bifurcation-theoretic origins of solutions and connections between them. A prominent example is in plane Couette flow; \citet{Reetz2019exact} constructed the first equilibrium solution underlying the self-organized oblique turbulent-laminar stripe pattern, and suggested that it emerges from the well-studied Nagata equilibrium \citep{Nagata1990}. Many other examples in convective systems can be found in \citet{Boronska2010_part1, Boronska2010_part2}, \citet{Reetz2020a, Reetz2020b} and \citet{Zheng2024part1, Zheng2024part2}.

\par The present work follows two previous numerical investigations \citep{Gao2018, Zheng2024part2}. \citet{Gao2018} have surveyed the flow regimes in a three-dimensional computational domain of size $[L_x,L_y,L_z]=[1,8,9]$, depicted in figure \ref{part3_VC_figure}, by systematically increasing the Rayleigh number from the onset of convection at $Ra=5707$ to $Ra\approx6300$ (with $Pr=0.71$ corresponding to air). \citet{Zheng2024part2} used the same domain size, constructed invariant solutions, and extended the upper limit to $Ra\approx6400$; a sequence of bifurcations was determined, and six equilibria and four time-periodic solutions were analysed in detail. In addition to these known solutions, we present here 33 new unstable invariant solutions, including seven equilibria and 26 periodic orbits, and we have extended the Rayleigh number range to $Ra\approx6650$. Even though the increase in $Ra$ in each paper may seem insignificant and negligible compared to fully turbulent convection, the emerging complexity of the bifurcation problem is indeed already overwhelming.

\par The new solutions that we will discuss are mainly found by the standard recurrent flow analysis which uses time-dependent simulations to locate states at which nearly-periodic solutions or near recurrences are detected, and uses them as initial conditions for Newton solving. By construction, then, the new solutions are embedded in the trajectories followed by the flow. The new periodic orbits not only capture important dynamics of the transitional flow, but also give hope that they may act as a basis to predict the statistical quantities of the dynamics \citep{Hopf1948}; these two points will be discussed in detail in \S \ref{part3_section_PP}. For recent analysis of statistical descriptions based on periodic orbits in two-dimensional Kolmogorov flow, see for instance \citet{Chandler2013, Cvitanovic2013, Lucas2015, Page2024}.

\par The rest of the manuscript is structured as follows. The numerical methods are summarized in \S \ref{part3_methods}. We discuss in \S \ref{part3FP} the new equilibria and in \S \ref{part3_UPO} the new periodic orbits, with a focus on their bifurcation scenarios. Section \ref{part3_section_PP} explores the dynamical relevance of the identified orbits and discusses a statistical description based on unstable orbits. The manuscript will conclude with future research directions in \S \ref{part3_conclusion}.

\section{System, computation of invariant solutions and symmetries}
\label{part3_methods}
\begin{figure}
    \centering
    \includegraphics[width=0.45\columnwidth]{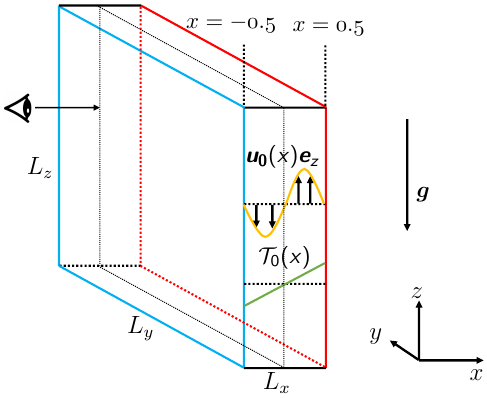}
    \captionsetup{font={footnotesize}}
    \captionsetup{width=13.5cm}
    \captionsetup{format=plain, justification=justified}
    \caption{\label{part3_VC_figure} Vertical convection cell with size $[L_x, L_y, L_z] = [1, 8, 9]$. The flow is bounded between two fixed walls at $x=\pm0.5$ at which the flow is heated and cooled respectively. We visualize the flow on the $y$-$z$ plane at $x=0$ (dotted), from left to right as indicated by the eye and arrow. The laminar velocity $\boldsymbol u_0(x) = \sqrt{Ra/Pr} (x/4 - x^3)/6 \:\boldsymbol e_z$ and temperature $\mathcal{T}_0(x) = x$ of this system are traced as an orange curve and a green line, respectively.}
\end{figure}

\par As the numerical methods used in the current research are exactly the same as those described in our precedent Part 1 and Part 2 papers, we refer readers to \citet{Zheng2024part1, Zheng2024part2} as well as the introduction of \citet{Zheng2025PhD} for detailed descriptions of the governing equations, laminar base solutions, boundary conditions, symmetries of the system, computation (including parametric continuation and linear stability analysis) of invariant solutions, as well as numerical visualizations. Here, we will succinctly summarize the key ingredients.

\par The Oberbeck–Boussinesq equations that govern our vertical convection system are
\begin{subequations}
\label{part3_appabc}
\begin{align}
    \dfrac{\partial \boldsymbol u}{\partial t} + (\boldsymbol u \cdot \nabla)\boldsymbol u &= -\nabla p + \left(\frac{Pr}{Ra}\right)^{1/2} \nabla^2 \boldsymbol u + \mathcal{T}\boldsymbol e_z, \\
    \dfrac{\partial \mathcal{T}}{\partial t} + (\boldsymbol u \cdot \nabla)\mathcal{T} &= \left(\frac{1}{Pr\: Ra}\right)^{1/2}\nabla^2 \mathcal{T}, \\
    \nabla \cdot \boldsymbol u&= 0.	
\end{align}
\end{subequations}
Equations \eqref{part3_appabc} with periodic boundary conditions (in $y$ and $z$), Dirichlet boundary conditions (in $x$), and a zero mean pressure gradient (in $y$ and $z$) integral constraint, are simulated numerically by using the Channelflow 2.0 code \citep{Gibson2019}. As in \citet{Zheng2024part2}, we use a computational domain of size $[L_x, L_y, L_z] = [1, 8, 9]$ and discretize it by $[N_x, N_y, N_z] = [31,96,96]$ Chebychev-Fourier–Fourier modes, see figure \ref{part3_VC_figure}. In addition to the grid resolution criterion discussed in \S 2.2 of \citet{Zheng2024part1}, we carried out a grid independence study by increasing $[N_x, N_y, N_z]$ to $ [41,136,136]$ modes; the results are described in \textit{Appendix} \ref{part3_appendix_grid}.

\par The symmetries of the system are:
\begin{subequations}
    \begin{eqnarray}
        &\pi_y[u,v,w,\mathcal{T}](x,y,z) \equiv [u,-v,w,\mathcal{T}](x, -y,z) \label{part3_sym_a},\\
        &\pi_{xz}[u,v,w,\mathcal{T}](x,y,z) \equiv [-u,v,-w,-\mathcal{T}](-x, y,-z) \label{part3_sym_b}, \\
        &\tau(\Delta y, \Delta z)[u,v,w,\mathcal{T}](x,y,z) \equiv [u,v,w,\mathcal{T}](x, y + \Delta y,z + \Delta z). \label{part3_sym_c}
    \end{eqnarray}
\end{subequations}
Definitions \eqref{part3_sym_a}--\eqref{part3_sym_c} stand for reflection in $y$, combined reflection of $x$, $z$ and temperature $\mathcal{T}$, and translation in $y$ and $z$, respectively. These symmetry operations generate the group $S \equiv \braket{\pi_y, \pi_{xz}, \tau(\Delta y, \Delta z)} \sim [O(2)]_y \times [O(2)]_{x,z}$, where $[O(2)]_y$ refers to reflections and translations in $y$, as in \eqref{part3_sym_a} and \eqref{part3_sym_c}, respectively, while $[O(2)]_{xz}$ refers to reflections in $(\mathcal{T},x,z)$ as in \eqref{part3_sym_b} and translations in $z$ as in \eqref{part3_sym_c}. The laminar base flow has symmetry $[O(2)]_y \times [O(2)]_{x,z}$, but symmetries are often sequentially broken at each bifurcation. We will discuss the symmetries of each solution in \S \ref{part3FP} and \S \ref{part3_UPO}. In addition, symmetries are also used as a tool to find invariant solutions in this work, because if solutions have such symmetries, they are usually less unstable in the constrained symmetry subspace, thus less difficult to find and converge.

\par Invariant solutions (equilibria and periodic orbits) are state vectors $\boldsymbol{x}^{*}(t)$ satisfying 
\begin{equation}
    \mathcal{G}(\boldsymbol{x}^{*})=\sigma \mathcal{F}^T(\boldsymbol{x}^{*}) - \boldsymbol{x}^{*} = 0,	
    \label{Part3_invariant_equation}
\end{equation}
where $\sigma$ is a symmetry operator and $\mathcal{F}^T$ is the time-evolution operator integrating \eqref{part3_appabc} from an initial state $\boldsymbol{x^*}$ over a finite time period $T$. $T$ is arbitrary for a steady solution, and is the period of a time-periodic solution. The periodic orbits that we will discuss in \S \ref{part3_UPO} are of three types. Periodic orbits (POs) are solutions which recur exactly after a period ($\sigma$ is the identity in \eqref{Part3_invariant_equation}). Relative periodic orbits (RPOs) are orbits whose shortest recurrence occurs for a non-trivial symmetry operation, e.g.\, $\sigma \equiv \tau(\Delta y, \Delta z)$. Pre-periodic orbits (PPOs) are RPOs which recur exactly after some finite number of periods, see e.g. \citet{cvitanovic2005chaos, Budanur2017}. (That is, PPOs are RPOs in which $\sigma$ does not contain translations by irrational multiples of $L_y$ or $L_z$.) In most of the later figures, we use $T$ to denote periods for PO, relative periods for RPO, and pre-periods for PPO. Whenever we discuss a period $T$ of an orbit in the text, we specify to which type of period it refers; in most cases, it is the shortest period for which \eqref{Part3_invariant_equation} holds for some $\sigma$.

\par Invariant solutions are computed by the shooting-based Newton method, with initial guesses generated by a systematic recurrent flow analysis. The success rate for converging to an invariant solution from one of these initial guesses is roughly $40\%$. These solutions are then parametrically continued in Rayleigh number to construct bifurcation diagrams. We define the temperature deviation $\theta \equiv \mathcal{T}-\mathcal{T}_0$ from the conductive state $\mathcal{T}_0$ shown in figure \ref{part3_VC_figure}. We use its $L_2$-norm $\lvert\lvert \theta \lvert\lvert_2$ to plot bifurcation diagrams throughout \S \ref{part3FP} and \S \ref{part3_UPO}. The linear stability of converged states is evaluated by the Arnoldi algorithm. All of the 33 new solution branches are linearly unstable; they will be shown as solid curves in bifurcation diagrams. The thermal energy input ($I$) due to buoyancy forces and the dissipation ($D$) due to viscosity, both averaged over the domain, are used for phase portrait visualizations; we refer readers to \S 2.3 of \citet{Reetz2020a} for formulas. The flows (and eigenvectors) are visualized via their temperature fields $\theta$ on the $y$-$z$ plane at $x=0$, see figure \ref{part3_VC_figure}. In order to avoid overcrowding the figures, we have omitted colour bars in all snapshots while insuring that all snapshots in a single figure share the same colour bar.

\section{Unstable equilibria}
\label{part3FP}
\begin{figure}
    \centering
    \includegraphics[width=\columnwidth]{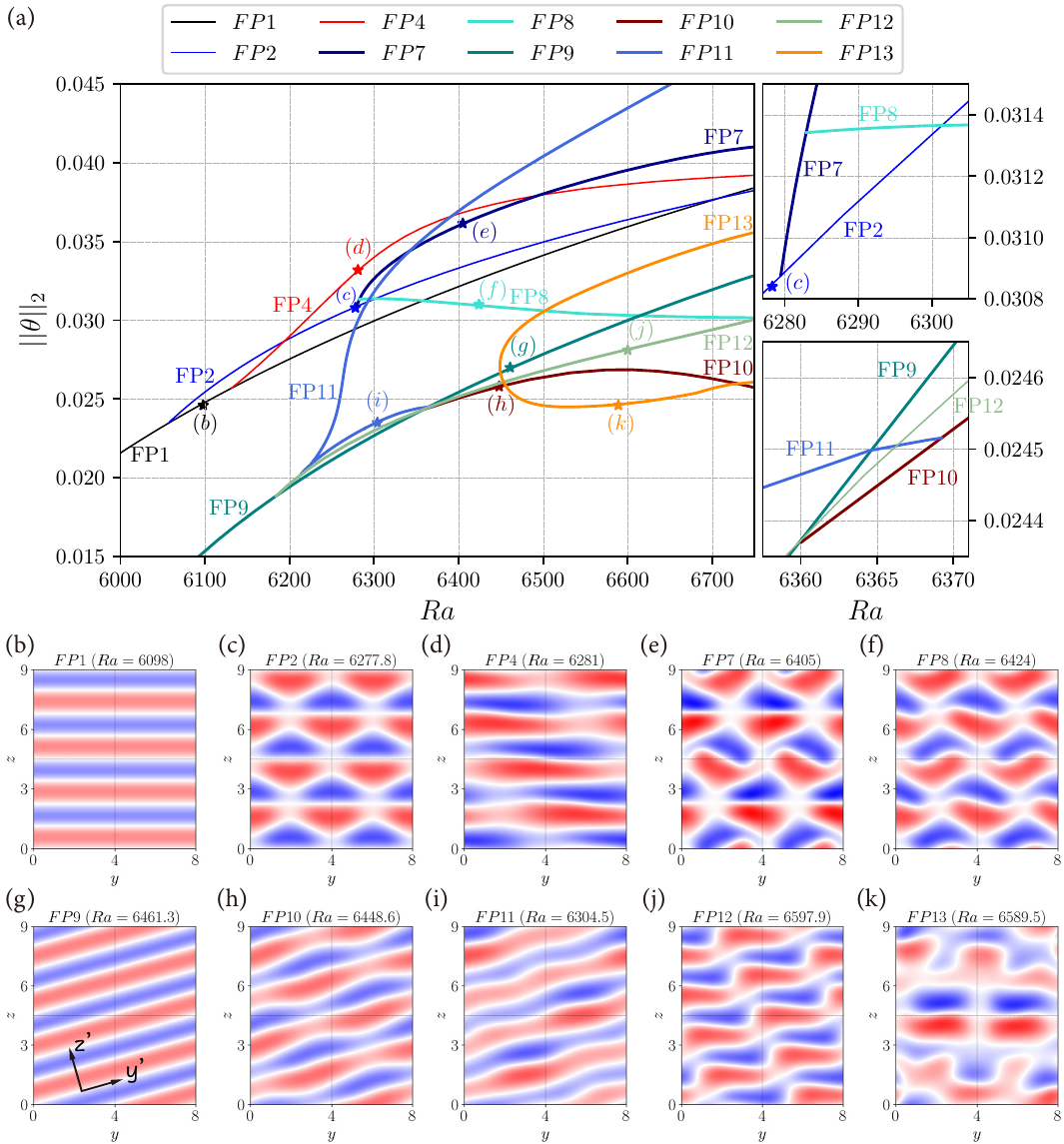}
    \captionsetup{font={footnotesize}}
    \captionsetup{width=13.5cm}
    \captionsetup{format=plain, justification=justified}
    \caption{\label{part3_BD-newFP} (a) Bifurcation diagram of equilibria and (b-k) flow structures visualized via the midplane temperature field. (b) FP1, (c) FP2 and (d) FP4 have been presented in \citet{Zheng2024part2} and are shown with thinner curves in (a). All branches shown are unstable, with the exception of FP1 for $Ra<6056$ and of FP2 for $6056<Ra<6058.5$. Two enlarged diagrams are shown on the right zooming in on the FP2$\rightarrow$FP7$\rightarrow$FP8 and FP9$\rightarrow$FP10$\rightarrow$FP11 bifurcations. (e) FP7 bifurcates from FP2 at $Ra=6279.5$; (f) FP8 bifurcates from FP7 at $Ra=6282.9$. (g) FP9 bifurcates from the unstable base state at $Ra=5941$; (h) FP10 bifurcates from FP9 at $Ra=6360$; (i) FP11 bifurcates from FP10 at $Ra=6369.2$ and undergoes a saddle--node bifurcation at $Ra=6213.5$; (j) FP12 bifurcates from FP9 at $Ra=6184$. (k) FP13 undergoes a saddle--node bifurcation at $Ra=6449$ and both upper and lower branches exist at least until $Ra=6800$. The intersection of the saddle--node bifurcation point of FP11 and FP13 with the nearby branches is due only to projection and does not represent a bifurcation.}
\end{figure}

\par In addition to the six equilibrium solutions (FP1--FP6) presented in \citet{Zheng2024part2}, we discuss here seven new unstable equilibria: FP7--FP13. These equilibria are relevant for the discussion on Hopf and global bifurcations of periodic orbits in \S \ref{part3_UPO}. A bifurcation diagram including all of the new steady states as well as FP1, FP2 and FP4 (FP3, FP5 and FP6 are not plotted to avoid clutter) is illustrated in figure \ref{part3_BD-newFP}(a). All of the equilibria are continued forward in Rayleigh number until at least $Ra=6750$. Note that many other branches of equilibria (and periodic orbits) exist, which we have not found, followed or shown on these diagrams.

\subsection{Equilibria FP7--FP8}
\par We begin our survey by briefly discussing FP1 and FP2. Equilibrium FP1 (2D rolls) is shown in figure \ref{part3_BD-newFP}(b) and contains four straight convection rolls of wavelength $\lambda_{\text{FP1}} =9/4=2.25$ whose axes are oriented in the $y$ direction. Equilibrium FP2 bifurcates from FP1 at $Ra = 6056$. Equilibrium FP2 is called wavy rolls in \citet{Gao2018} and diamond rolls in \citet{Zheng2024part2}, and is shown in figure \ref{part3_BD-newFP}(c). (The list of generators for FP2 in \eqref{part3_sym_FP7FP8} omits $\tau(0, L_z/2)$, contained in (3.1) of \citet{Zheng2024part2}, because it can be produced by the other generators and is hence redundant.) Equilibrium FP7, shown in figure \ref{part3_BD-newFP}(e), bifurcates from FP2 at $Ra=6279.5$ in a supercritical pitchfork bifurcation, in which the $\pi_y$ reflection and four-fold translation (along both diagonals) symmetries are broken. Equilibrium FP8, shown in figure \ref{part3_BD-newFP}(f), bifurcates from FP7 in a supercritical pitchfork bifurcation at $Ra=6282.9$, in which the $\pi_y\tau(0, L_z/2)$ symmetry is broken. Equilibrium FP8 gives rise to PO23 and PO24 in two Hopf bifurcations, see \S \ref{part3_sym_ref}. The symmetry groups of FP1, FP2, FP7 and FP8 are
\begin{equation}
    \begin{array}{lll}
        \text{FP1:} \; & \braket{\pi_y, \tau(\Delta y,0),\pi_{xz},\tau(0,L_z/4)} & \sim [O(2)]_y \times [D_4]_{xz};\\
        \text{FP2:} \; & \braket{\tau(L_y/2,0), \pi_{xz}, \pi_y, \tau(L_y/4, -L_z/4)} & \sim D_2 \times D_4; \\ 
        \text{FP7:} \; & \braket{\tau(L_y/2,0), \pi_{y}\pi_{xz}, \pi_y\tau(0, L_z/2)} & \sim D_2 \times Z_2; \\ 
        \text{FP8:} \; & \braket{\tau(L_y/2,0), \pi_{y}\pi_{xz}} & \sim D_2.
    \end{array}
\label{part3_sym_FP7FP8}
\end{equation} 

\subsection{Equilibria FP9--FP12} 
\label{part3_FP9_12}
\par Equilibrium FP9, shown in figure \ref{part3_BD-newFP}(g), bifurcates from the homogeneous unstable base flow at $Ra=5941$ (not shown in figure \ref{part3_BD-newFP}a) in a supercritical pitchfork bifurcation. Equilibrium FP9 has four pairs of oblique but straight convection rolls of wavelength $\lambda_{\text{FP9}} =2L_z/\sqrt{L_z^2/16+L_y^2} \approx2.166$ each, in the direction perpendicular to the rolls. The oblique angle with respect to the $y$-direction is $\gamma=\arctan(0.25L_z/L_y)\approx 15.7\degree$. Because FP9--FP12 all share this oblique orientation, we introduce tilted coordinates
\begin{align}
\left(\begin{array}{c} y^\prime \\ z^\prime \end{array}\right) 
= \left(\begin{array}{rr} 
\cos{\gamma} & \sin{\gamma} \\
-\sin{\gamma} & \cos{\gamma}
\end{array}\right)
\left(\begin{array}{c} y\\z \end{array}\right),
\end{align}
which are drawn in figure \ref{part3_BD-newFP}(g). In the tilted coordinates, we consider a virtual computational domain having length $L_z^\prime=4\lambda_{\text{FP9}}\approx8.664$ in $z^\prime$ and $L_y^\prime \approx 15$ in $y^\prime$. (The length $L_y^\prime \approx 15$ is three times the wavelength corresponding to the prominent structure along $y^\prime$ in FP10 and FP11, and four times the wavelength corresponding to the wavy structure in FP12. Introducing this length will be convenient for the description of symmetry groups.) In this tilted domain, FP9 has $O(2)$ symmetry in $y^\prime$ and $D_4$ symmetry in $xz^\prime$; see \eqref{part3_eqn_sym_FP9_10_11_12}.

\par Equilibrium FP10, shown in figure \ref{part3_BD-newFP}(h), bifurcates from FP9 at $Ra=6360$ in a supercritical pitchfork bifurcation. In this bifurcation, the $O(2)$ symmetry of FP9 along $y^\prime$ is broken and succeeded by a discrete (two-fold) translation. In $z^\prime$, the four-fold translation must now be combined with a discrete (four-fold) translation in $y^\prime$ in order to remain a symmetry of the flow. Finally, the two independent reflection symmetries are replaced by a single combined reflection $\pi_{y^\prime}\pi_{xz^\prime}$.

\par Equilibrium FP11, shown in figure \ref{part3_BD-newFP}(i), bifurcates from FP10 at $Ra=6369.2$ in a subcritical pitchfork bifurcation. Equilibrium FP11 then undergoes a saddle--node bifurcation at $Ra=6213.5$ and continues to exist at least until $Ra=7000$. In going from FP10 to FP11, the spatial periodicity along $y^\prime$ changes from $L_y^\prime/2$ to $L_y^\prime$, while other symmetries are retained. Equilibrium FP11 gives rise to PO14 in a Hopf bifurcation, see \S \ref{part3_UPO14} in which FP9 (and FP12 below) will also be relevant.

\par Equilibrium FP12, shown in figure \ref{part3_BD-newFP}(j), also bifurcates from FP9, in a supercritical pitchfork bifurcation at $Ra=6184$. In $y^\prime$, the $O(2)$ symmetry of FP9 is succeeded by a four-fold translation, the four-fold translation in $z^\prime$ is retained, and the two reflection symmetries are replaced by the single combined reflection $\pi_{y^\prime}\pi_{xz^\prime}$.

\par The symmetry groups of FP9--FP12 are
\begin{equation}
    \begin{array}{lll}
     \text{FP9:} \; & \braket{\pi_{y^\prime}, \tau(\Delta y^\prime, 0),\pi_{xz^\prime}, \tau(0,L_z^\prime/4)} & \sim [O(2)]_{y^\prime} \times [D_4]_{xz^\prime}; \\
     \text{FP10:} \; & \braket{\pi_{y^\prime}\pi_{xz^\prime}, \tau(L_y^\prime/2, 0), \tau(L_y^\prime/4, L_z^\prime/4)} & \sim [D_2]_{y^\prime} \times [Z_4]_{xz^\prime}; \\
     \text{FP11:} \; & \braket{\pi_{y^\prime}\pi_{xz^\prime}, \tau(L_y^\prime/4, L_z^\prime/4)} & \sim [Z_2]_{y^\prime} \times [Z_4]_{xz^\prime}; \\
     \text{FP12:} \; & \braket{\pi_{y^\prime}\pi_{xz^\prime}, \tau(L_y^\prime/4, 0), \tau(0,L_z^\prime/4)} & \sim [D_4]_{y^\prime} \times [Z_4]_{xz^\prime};
    \end{array}
\label{part3_eqn_sym_FP9_10_11_12}
\end{equation}

\subsection{Equilibrium FP13}
\par Equilibrium FP13 is shown in figure \ref{part3_BD-newFP}(k) and exists beyond $Ra=6800$, where we stopped the continuation; its bifurcation-theoretic origin remains unclear. In the Rayleigh number range that we consider, FP13 undergoes one saddle--node bifurcation at $Ra=6449$. Equilibrium FP13 is relevant for PO23 in \S \ref{part3_UPO23}, and its symmetry group is
\begin{equation}
    \begin{array}{lll}
        \text{FP13:} \; \braket{\pi_{y}\pi_{xz}} & \sim Z_2. \\
    \end{array}
\end{equation}

\section{Unstable periodic orbits}
\label{part3_UPO}
\begin{figure}
    \centering
    \includegraphics[width=\columnwidth]{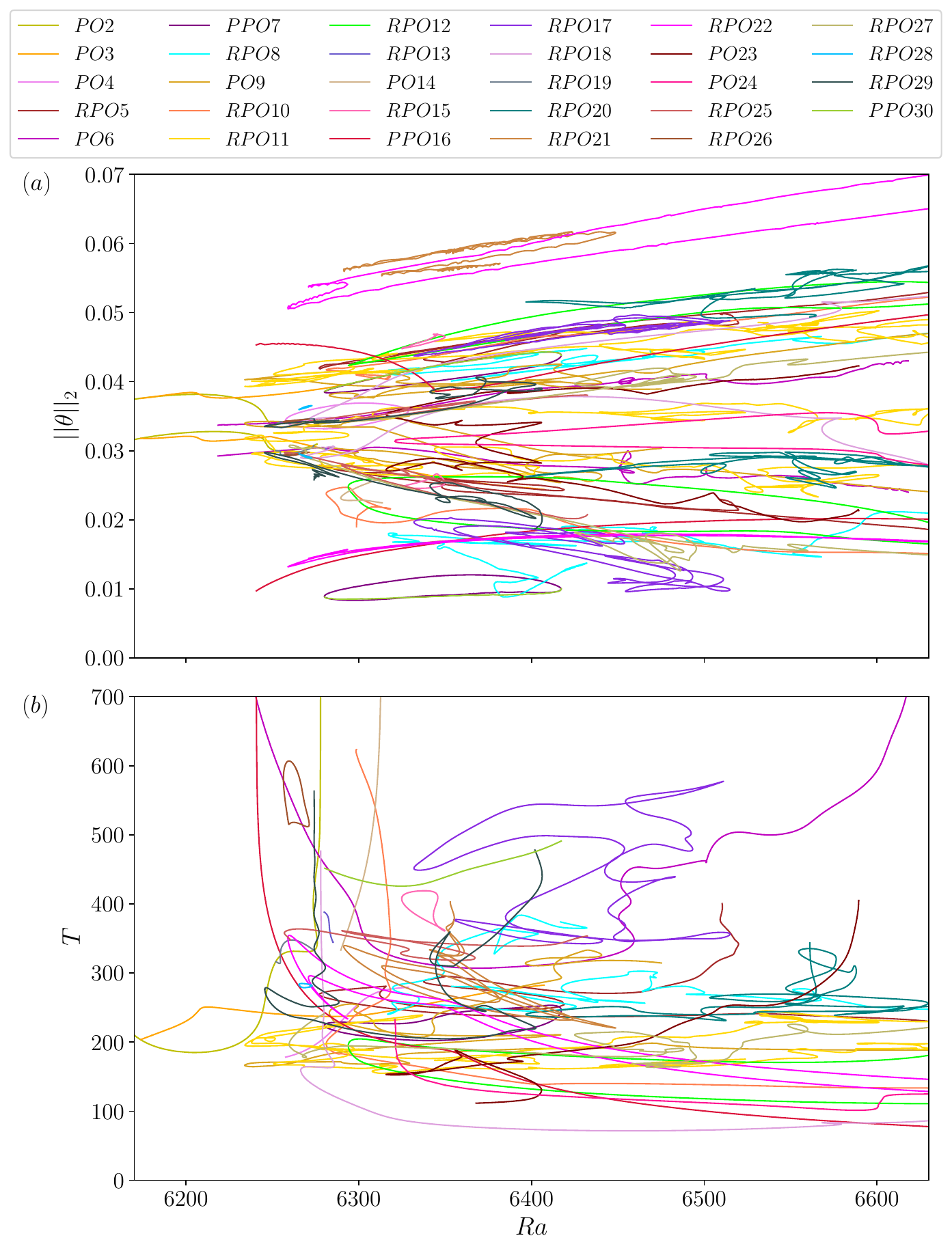}
    \captionsetup{font={footnotesize}}
    \captionsetup{width=13.5cm}
    \captionsetup{format=plain, justification=justified}
    \caption{\label{part3_BD-newPO} Temperature norms (a) and 
    periods (b) of periodic orbits. Abbreviations PO, RPO and PPO stand for periodic orbit, relative periodic orbit and pre-periodic orbit, respectively. Orbits PO2--PO4 are discussed in detail in \citet{Zheng2024part2}. In (a), for each orbit, we show two curves, the maximum and minimum of $\lvert\lvert \theta \lvert\lvert_2$ along an orbit. All of RPO5--PPO30 are linearly unstable. The upper limit of (b) is set to $T=700$, even though some orbits are continued to higher period. The bifurcation scenarios include Hopf, pitchfork, saddle--node, period-doubling, period-halving, global homoclinic/heteroclinic bifurcations and isolas. For more clarity, bifurcation diagrams for selected sets of orbits will be shown in figures \ref{part3_sepa_BD_PO13_15_26_28}, \ref{part3_sepa_BD_PO17_18_27}, \ref{part3_sepa_BD_PO19_25}, \ref{part3_sepa_BD_PO6_10_14_16_29}, \ref{part3_sepa_BD_PO5_7_8_9_30}, \ref{part3_sepa_BD_PO11_12_20} and \ref{part3_sepa_BD_PO21_22_23_24}. The apparent lack of smoothness in some $\lvert\lvert \theta \lvert\lvert_2$ curves corresponds to the overtaking of one temporal maximum or minimum of $\lvert\lvert \theta \lvert\lvert_2$ by another as $Ra$ is varied.}
\end{figure}

\begin{table}
    \centering
    \begin{tabular}{cccccc}
        Spatial symmetry & Periodic orbit & Bifurcations & Discussed in & Shown in figure(s) \\ [0.2cm]
        $\braket{\pi_{y}\pi_{xz}, \tau(L_y/4, L_z/4)}$ & PO1 & H, SN, GB & \citet{Zheng2024part2} & 5, 6, 7, 8, 15 \\
        $\braket{\pi_{y}\pi_{xz}, \tau(L_y/4, L_z/4)}$ & PO2 & PD, SN, GB & \citet{Zheng2024part2} & 5, 9, 10, 11, 15 \\
        $\braket{\tau(L_y/2, L_z/2)}$ & PO3 & PF & \citet{Zheng2024part2} & 5, 13, 15 \\
        $\braket{\pi_{y}, \pi_{xz}\tau(L_y/2, 0)}$ & PO4 & H & \citet{Zheng2024part2} & 5, 14, 15 \\ [0.2cm]
        $\braket{\pi_{y}\pi_{xz}, \tau(L_y/4, L_z/4)}$ & RPO13 & PD & \S \ref{part3_UPO13} & \ref{part3_sepa_BD_PO13_15_26_28}\\
        $\braket{\pi_{y}\pi_{xz}, \tau(L_y/4, L_z/4)}$ & RPO15 & SN, isola & \S \ref{part3_isola_RPO15-17-26-28} & \ref{part3_sepa_BD_PO13_15_26_28}\\
        $\braket{\pi_{y}\pi_{xz}, \tau(L_y/4, L_z/4)}$ & RPO17 & SN, isola & \S \ref{part3_isola_RPO15-17-26-28} & \ref{part3_sepa_BD_PO17_18_27}\\
        $\braket{\pi_{y}\pi_{xz}, \tau(L_y/4, L_z/4)}$ & RPO26 & SN, isola & \S \ref{part3_isola_RPO15-17-26-28} & \ref{part3_sepa_BD_PO13_15_26_28}\\
        $\braket{\pi_{y}\pi_{xz}, \tau(L_y/4, L_z/4)}$ & RPO28 & SN, isola & \S \ref{part3_isola_RPO15-17-26-28} & \ref{part3_sepa_BD_PO13_15_26_28}\\
        $\braket{\pi_{y}\pi_{xz}, \tau(L_y/4, L_z/4)}$ & RPO18 & SN, GB & \S \ref{part3_UPO18} & \ref{part3_sepa_BD_PO17_18_27}, \ref{part3_PO18_series_snapshots}\\
        $\braket{\pi_{y}\pi_{xz}, \tau(L_y/4, L_z/4)}$ & RPO19 & PD, SN & \S \ref{part3_UPO19} & \ref{part3_sepa_BD_PO19_25} \\ [0.2cm]
        $\braket{\pi_{y}\pi_{xz}, \tau(L_y/2, L_z/2)}$ & PO6 & SN, GB & \S \ref{part3_UPO6} & \ref{part3_sepa_BD_PO6_10_14_16_29}, \ref{part3_PO6hetero}, \ref{part3_PO6robustness} \\
        $\braket{\pi_{y}\pi_{xz}, \tau(L_y/2, L_z/2)}$ & PPO7 & SN, isola & \S \ref{part3_UPO7} & \ref{part3_sepa_BD_PO5_7_8_9_30}, \ref{part3_PO7_series_snapshots} \\
        $\braket{\pi_{y}\pi_{xz}, \tau(L_y/2, L_z/2)}$ & RPO10 & SN, GB & \S \ref{part3_UPO10} & \ref{part3_sepa_BD_PO6_10_14_16_29}, \ref{part3_PO10_series_snapshots}\\
        $\braket{\pi_{y}\pi_{xz}, \tau(L_y/2, L_z/2)}$ & RPO25 & PF, SN, PH & \S \ref{part3_UPO25} & \ref{part3_sepa_BD_PO19_25}\\
        $\braket{\pi_{y}\pi_{xz}, \tau(L_y/2, L_z/2)}$ & RPO27 & PF, SN & \S \ref{part3_UPO27} & \ref{part3_sepa_BD_PO17_18_27}\\
        $\braket{\pi_{y}\pi_{xz}, \tau(L_y/2, L_z/2)}$ & RPO29 & SN, GB & \S \ref{part3_UPO29} & \ref{part3_sepa_BD_PO6_10_14_16_29}, \ref{part3_PO29_series_period}\\
        $\braket{\pi_{y}\pi_{xz}, \tau(L_y/2, L_z/2)}$ & PPO30 & PD & \S \ref{part3_UPO30} & \ref{part3_sepa_BD_PO5_7_8_9_30}\\ [0.2cm]
        $\braket{\tau(L_y/4, L_z/4)}$ & RPO12 & SN, GB & \S \ref{part3_UPO12} & \ref{part3_sepa_BD_PO11_12_20}, \ref{part3_PO20_series_snapshots} \\
        $\braket{\tau(L_y/4, L_z/4)}$ & RPO20 & SN, GB & \S \ref{part3_UPO20} & \ref{part3_sepa_BD_PO11_12_20}, \ref{part3_PO20_series_snapshots} \\ [0.2cm]
        $\braket{\tau(L_y/2, L_z/2)}$ & RPO5 & SN, GB & \S \ref{part3_UPO5} & \ref{part3_sepa_BD_PO5_7_8_9_30}, \ref{part3_PO5PO8_series_snapshots}\\
        $\braket{\tau(L_y/2, L_z/2)}$ & RPO8 & SN & \S \ref{part3_UPO8} & \ref{part3_sepa_BD_PO5_7_8_9_30}, \ref{part3_PO5PO8_series_snapshots}\\ [0.2cm]
        $\braket{\pi_{y}\pi_{xz}, \tau(L_y/3, L_z/3)}$ & PO14 & H, GB & \S \ref{part3_UPO14} & \ref{part3_sepa_BD_PO6_10_14_16_29}, \ref{part3_PO14_series_snapshots}\\ [0.2cm]
        $\braket{\pi_{y}\pi_{xz}, \tau(L_y/5, L_z/5)}$ & PPO16 & GB & \S \ref{part3_UPO16} & \ref{part3_sepa_BD_PO6_10_14_16_29}, \ref{part3_PO16_series_snapshots}\\
        $\braket{\pi_{y}\pi_{xz}, \tau(L_y/5, L_z/5)}$ & RPO21 & SN & \S \ref{part3_UPO21_22} & \ref{part3_sepa_BD_PO21_22_23_24}\\
        $\braket{\pi_{y}\pi_{xz}, \tau(L_y/5, L_z/5)}$ & RPO22 & SN & \S \ref{part3_UPO21_22} & \ref{part3_sepa_BD_PO21_22_23_24}\\[0.2cm]
        $\braket{\pi_{y}\pi_{xz}}$ & PO23 & H, SN, GB & \S \ref{part3_UPO23} & \ref{part3_sepa_BD_PO21_22_23_24}, \ref{part3_PO23_series_snapshots}\\
        $\braket{\pi_{y}\pi_{xz}}$ & PO24 & H & \S \ref{part3_UPO24} & \ref{part3_sepa_BD_PO21_22_23_24}\\ [0.2cm]
        No spatial symmetry & PO9 & SN & \S \ref{part3_UPO9} & \ref{part3_sepa_BD_PO5_7_8_9_30}, \ref{part3_PO9_series_snapshots} \\
        No spatial symmetry & RPO11 & SN & \S \ref{part3_UPO11} & \ref{part3_sepa_BD_PO11_12_20}, \ref{part3_PO11_series_snapshots} \\
    \end{tabular}
    \captionsetup{font={footnotesize}}
    \captionsetup{width=13.5cm}
    \captionsetup{format=plain, justification=justified}
    \caption{Summary of spatial symmetries and bifurcation scenarios of 30 periodic orbits found in domain $[L_x,L_y,L_z]=[1,8,9]$, with PO1--PO4 discussed in \citet{Zheng2024part2}. PF, SN, PD, PH, H and GB are abbreviations for pitchfork, saddle--node, period-doubling, period-halving, Hopf and global bifurcations.}
    \label{part3_summary_UPO}
\end{table}

\par In this section we discuss 26 newly identified unstable periodic orbits RPO5--PPO30. Our naming convention is that the letters (e.g. R and P) describe the type of orbit, while the numbers (e.g. 5 and 30) identify different states based on the sequential order in which the orbits were found. Figure \ref{part3_BD-newPO} includes all of the periodic orbits PO2--PPO30 (PO1--PO4 are discussed in \cite{Zheng2024part2}) only to give an impression of the complexity of the full bifurcation diagram. To explain the various branches and scenarios, separated bifurcation diagrams for smaller groups of periodic orbits will be shown in figures \ref{part3_sepa_BD_PO13_15_26_28}, \ref{part3_sepa_BD_PO17_18_27}, \ref{part3_sepa_BD_PO19_25}, \ref{part3_sepa_BD_PO6_10_14_16_29}, \ref{part3_sepa_BD_PO5_7_8_9_30}, \ref{part3_sepa_BD_PO11_12_20} and \ref{part3_sepa_BD_PO21_22_23_24}. Given the complexity of all of the bifurcation diagrams, we recommend reading each diagram by first focusing on the plot of the temporal period and then comparing it with that of the temperature norm. The reason is that the quantity $\lvert\lvert \theta \lvert\lvert_2$ might sometimes be close for multiple orbits and for one orbit along the branch (due to saddle--node bifurcations), but the periods of the orbits are more distinct and thus lead to a better understanding of the bifurcation scenarios. In this work, we focus on Rayleigh numbers up to $Ra\approx6650$, thus $\sim$$16.5\%$ above the onset of convection, even though some orbits can be continued to much higher Rayleigh numbers. We do not discuss if and how their branches end there.

\par The bifurcation scenarios explored include Hopf, pitchfork, saddle--node, period-doubling, period-halving, global homoclinic/heteroclinic bifurcations and isolas. Given the large number of orbits that we will discuss, this section is organized in terms of the symmetries of the orbits. The eight subsections below will discuss orbits identified in the following symmetry subspaces: four-fold translation along the domain diagonal with a non-commuting reflection: $\braket{\pi_{y}\pi_{xz}, \tau(L_y/4, L_z/4)}$ in \S \ref{part3_sym_ref_4fold}; two-fold translation with a commuting reflection: $\braket{\pi_{y}\pi_{xz}, \tau(L_y/2, L_z/2)}$ in \S \ref{part3_sym_ref_2fold}; four-fold translation: $\braket{\tau(L_y/4, L_z/4)}$ in \S \ref{part3_sym_4fold}; two-fold translation: $\braket{\tau(L_y/2, L_z/2)}$ in \S \ref{part3_sym_2fold}; three-fold translation with a non-commuting reflection: $\braket{\pi_{y}\pi_{xz}, \tau(L_y/3, L_z/3)}$ in \S \ref{part3_sym_ref_3fold}; five-fold translation with a non-commuting reflection: $\braket{\pi_{y}\pi_{xz},\tau(L_y/5, L_z/5)}$ in \S \ref{part3_sym_ref_5fold}; and single reflection: $\braket{\pi_{y}\pi_{xz}}$ in \S \ref{part3_sym_ref}. Orbits without any spatial symmetry will be presented in \S \ref{part3_sym_no_sym}. Table \ref{part3_summary_UPO} provides a summary of these solutions in terms of their symmetries, bifurcation scenarios, section of coverage and figures in which they are shown.

\subsection{Symmetry subspace: reflection with four-fold translation}
\label{part3_sym_ref_4fold}
\par Seven orbits identified in the symmetry subspace $\braket{\pi_{y}\pi_{xz}, \tau(L_y/4, L_z/4)} \sim D_4$ will be discussed in this subsection. Due to this imposed symmetry constraint, the dynamics of these seven orbits all have a diagonal orientation and consist of diagonal excursions from more aligned states. We only show snapshots of RPO18 (figure \ref{part3_PO18_series_snapshots}) for illustration. 

\subsubsection{Orbit RPO13: period-doubling bifurcations}
\label{part3_UPO13}
\begin{figure}
    \centering
    \includegraphics[width=\columnwidth]{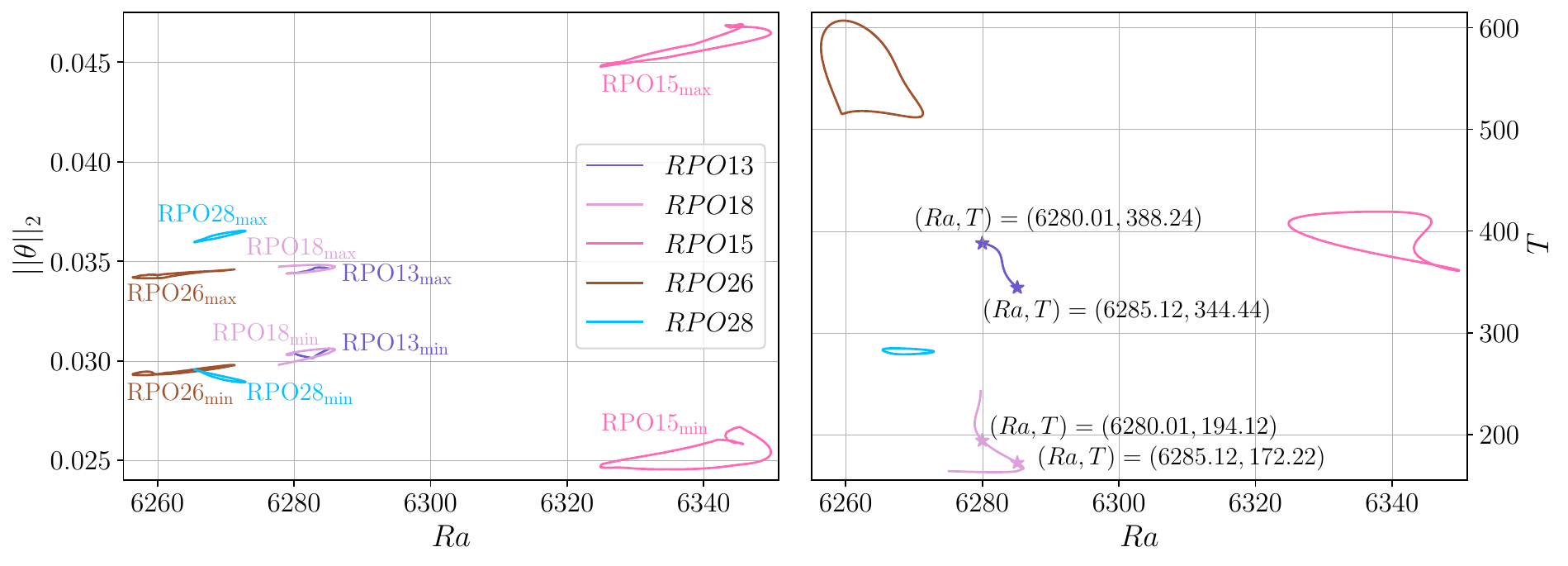}
    \captionsetup{font={footnotesize}}
    \captionsetup{width=13.5cm}
    \captionsetup{format=plain, justification=justified}  \caption{\label{part3_sepa_BD_PO13_15_26_28} Temperature norms (left) and periods (right) of RPO13, RPO15, RPO26 and RPO28. Branch RPO13 bifurcates from and terminates on RPO18 (which is shown more completely in figure \ref{part3_sepa_BD_PO17_18_27}) in two period-doubling bifurcations. The bifurcation points are indicated by stars on the right plot. Branches RPO15, RPO26 and RPO28 begin and terminate at saddle--node bifurcations and form isolas.}
\end{figure}

\par Orbit RPO13 was found at $Ra=6285$. Forward and backward continuation in Rayleigh number reveal that RPO13 bifurcates from and ends at RPO18 in two period-doubling bifurcations, at $Ra=6280.01$ and $6285.12$, as indicated by the stars in figure \ref{part3_sepa_BD_PO13_15_26_28}. (Orbit RPO18 will be discussed in \S \ref{part3_UPO18}.)

\subsubsection{Orbits RPO15, RPO17, RPO26 and RPO28: saddle--node bifurcations and isolas}
\label{part3_isola_RPO15-17-26-28}
\par We identify four isolas in this symmetry subspace. As the name implies, they do not bifurcate from any other states but only undergo several saddle--node bifurcations to turn back in $Ra$, as evidenced by figure \ref{part3_sepa_BD_PO13_15_26_28} for RPO15, RPO26, RPO28, and by figure \ref{part3_sepa_BD_PO17_18_27} for RPO17. One more isola with slightly different symmetry will be discussed in \S \ref{part3_UPO7}.

\subsubsection{Orbit RPO18: saddle--node and global bifurcations}
\label{part3_UPO18}

\begin{figure}
    \centering
    \includegraphics[width=\columnwidth]{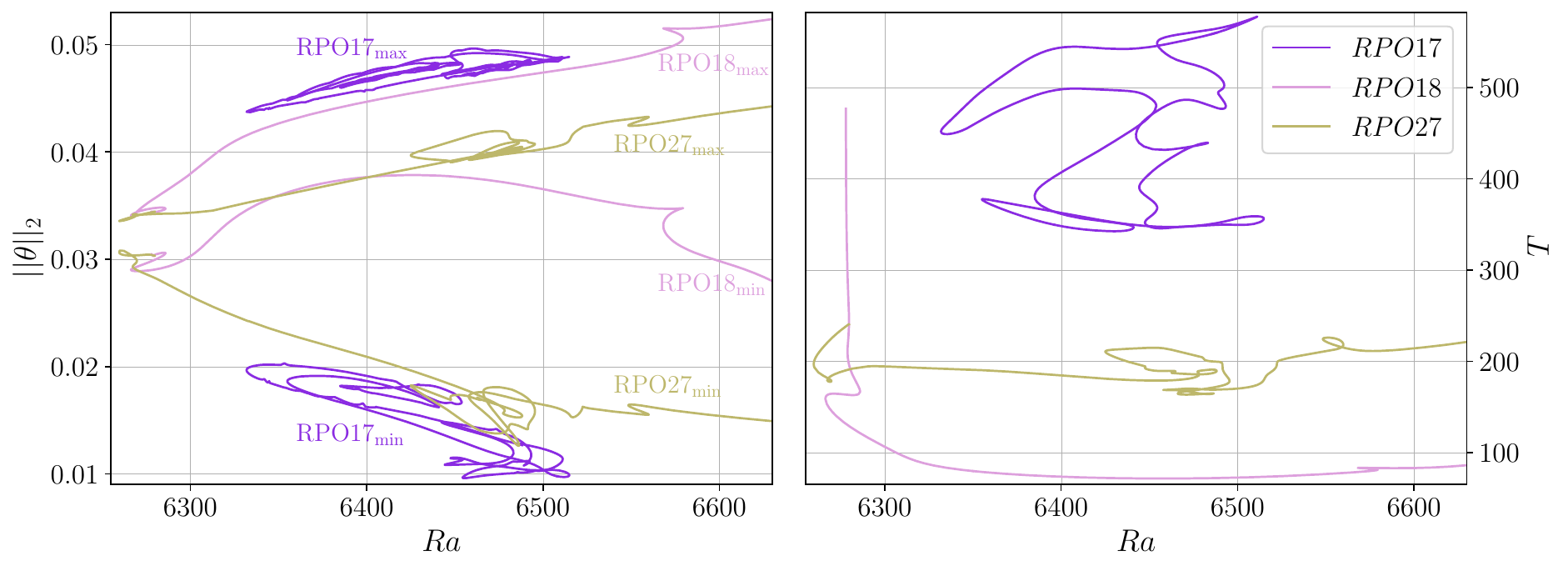}
    \captionsetup{font={footnotesize}}
    \captionsetup{width=13.5cm}
    \captionsetup{format=plain, justification=justified}
    \caption{\label{part3_sepa_BD_PO17_18_27} Temperature norms (left) and periods (right) of RPO17, RPO18 and RPO27. The RPO17 branch forms an isola. Orbit RPO18 bifurcates from FP2 in a global homoclinic bifurcation at $Ra=6277.96$ and continues to exist up to at least $Ra=6686$. Orbit RPO27 is generated from RPO18 in a pitchfork bifurcation at $Ra=6279.7$ and continues to exist up to at least $Ra=6650$.}
\end{figure}

\begin{figure}
    \centering
    \includegraphics[width=\columnwidth]{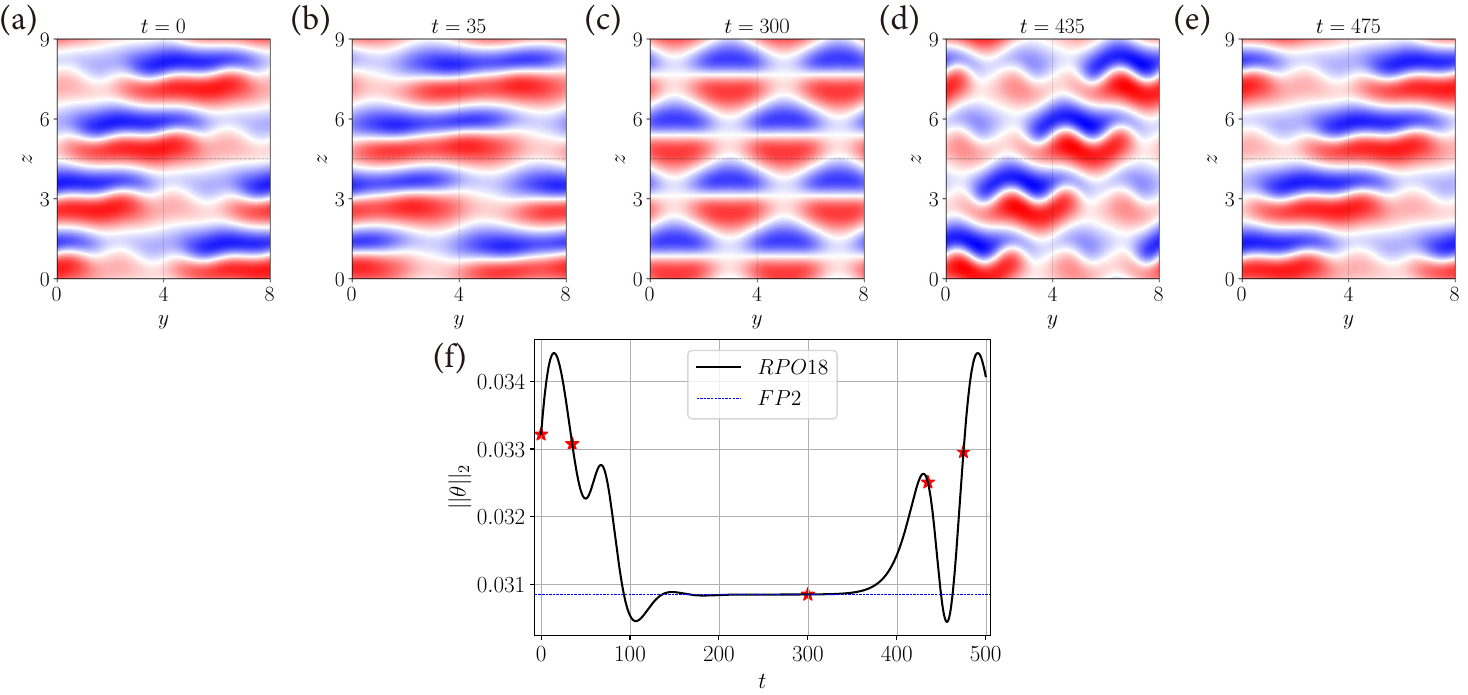}
    \captionsetup{font={footnotesize}}
    \captionsetup{width=13.5cm}
    \captionsetup{format=plain, justification=justified}
    \caption{\label{part3_PO18_series_snapshots} Dynamics of RPO18 at $Ra=6277.958$ (close to the global bifurcation point) with relative period $T=476.31$. (a--e) Snapshots of the midplane temperature field. (f) Time series from DNS. The five red stars indicate the moments at which the snapshots (a)--(e) are taken.}
\end{figure}

\par We found RPO18 at $Ra=6350$ and continued it forwards up to $Ra=6686$. Continuing backwards, RPO18 undergoes several saddle--node bifurcations and finally terminates in a global bifurcation by meeting FP2. The dynamics of RPO18 at $Ra=6277.958$, the lowest Rayleigh number that we have reached, is shown in figures \ref{part3_PO18_series_snapshots}(a-e). Orbit RPO18 resembles PO2, as presented in \citet{Zheng2024part2}: it has a clear oblique orientation. The global bifurcation is evidenced by the plateau in the time series in figure \ref{part3_PO18_series_snapshots}(f) and the divergence of its period in figure \ref{part3_sepa_BD_PO17_18_27}.

\par We have determined the eigenvector along which RPO18 approaches and escapes from FP2 by computing the leading eigenvalues of FP2 at $Ra=6277.958$ within the symmetry subspace $\braket{\pi_{y}\pi_{xz}, \tau(L_y/4, L_z/4)}$, as was done in \citet{Zheng2024part2}. We find that RPO18 escapes from FP2 along eigendirection $e_1$ associated with eigenvalue $\lambda_1=0.031285$ and approaches FP2 via eigendirection $e_2$ associated with eigenvalue $\lambda_2=-0.0138$. These eigenvectors turn out to be the same or symmetrically-related versions of those which are responsible for PO2 to escape from and to approach FP2. We do not show these to avoid repetition and refer readers to figures 9 and 10 of \citet{Zheng2024part2} for details. Interestingly, the global bifurcations of PO2 and RPO18 occur at almost the same Rayleigh number, and we will see in \S \ref{part3_UPO6} and \S \ref{part3_UPO29} that FP2 gives rise to other global heteroclinic and homoclinic bifurcations.

\subsubsection{Orbit RPO19: period-doubling and saddle--node bifurcations}
\label{part3_UPO19}
\begin{figure}
    \centering
    \includegraphics[width=\columnwidth]{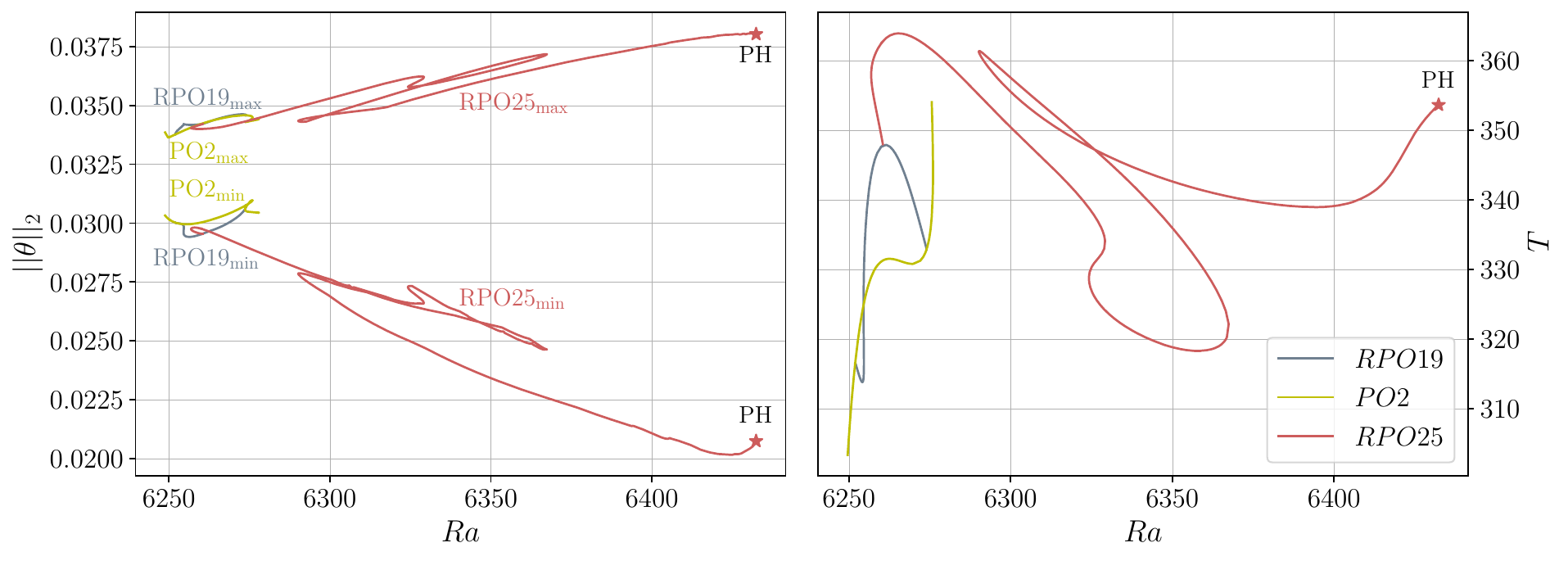}
    \captionsetup{font={footnotesize}}
    \captionsetup{width=13.5cm}
    \captionsetup{format=plain, justification=justified}
    \caption{\label{part3_sepa_BD_PO19_25} Temperature norms (left) and periods (right) of PO2, RPO19 and RPO25. Branch RPO19 bifurcates from and ends on PO2 (discussed in \citet{Zheng2024part2}) at $Ra=6252$ and $Ra=6274$ in two period-doubling bifurcations. Branch RPO25 bifurcates from RPO19 at $Ra=6260.5$ in a pitchfork bifurcation, undergoes saddle--node bifurcations, and terminates in a period-halving bifurcation (marked by PH) on another branch that is not shown or studied in this paper.}
\end{figure}

\par As shown in figure \ref{part3_sepa_BD_PO19_25}, RPO19 bifurcates from and terminates on PO2 in two period-doubling bifurcations. As discussed in \S 4.2 of \citet{Zheng2024part2}, PO2 has a spatio-temporal symmetry and contains two pre-periodic orbits, each of half period of PO2. The periods of PO2 that we show in figure \ref{part3_sepa_BD_PO19_25} (and figure \ref{part3_BD-newPO}b) correspond to full periods or twice the pre-periods. Orbit RPO19 also undergoes two saddle--node bifurcations at $Ra\approx 6254.6$, which are difficult to see in the figure.

\subsection{Symmetry subspace: reflection with two-fold translation}
\label{part3_sym_ref_2fold}
\par Seven orbits identified in the symmetry subspace $\braket{\pi_{y}\pi_{xz}, \tau(L_y/2, L_z/2)} \sim D_2$ will be discussed in this subsection. Similarly to \S \ref{part3_sym_ref_4fold}, this imposed symmetry leads the dynamics of most of the orbits to acquire a diagonal orientation. We will show snapshots of PO6 (diagonal), PPO7 (non-diagonal) and RPO10 (diagonal), for illustration. 

\subsubsection{Orbit PO6: saddle--node and global bifurcations}
\label{part3_UPO6}

\begin{figure}
    \centering
    \includegraphics[width=\columnwidth]{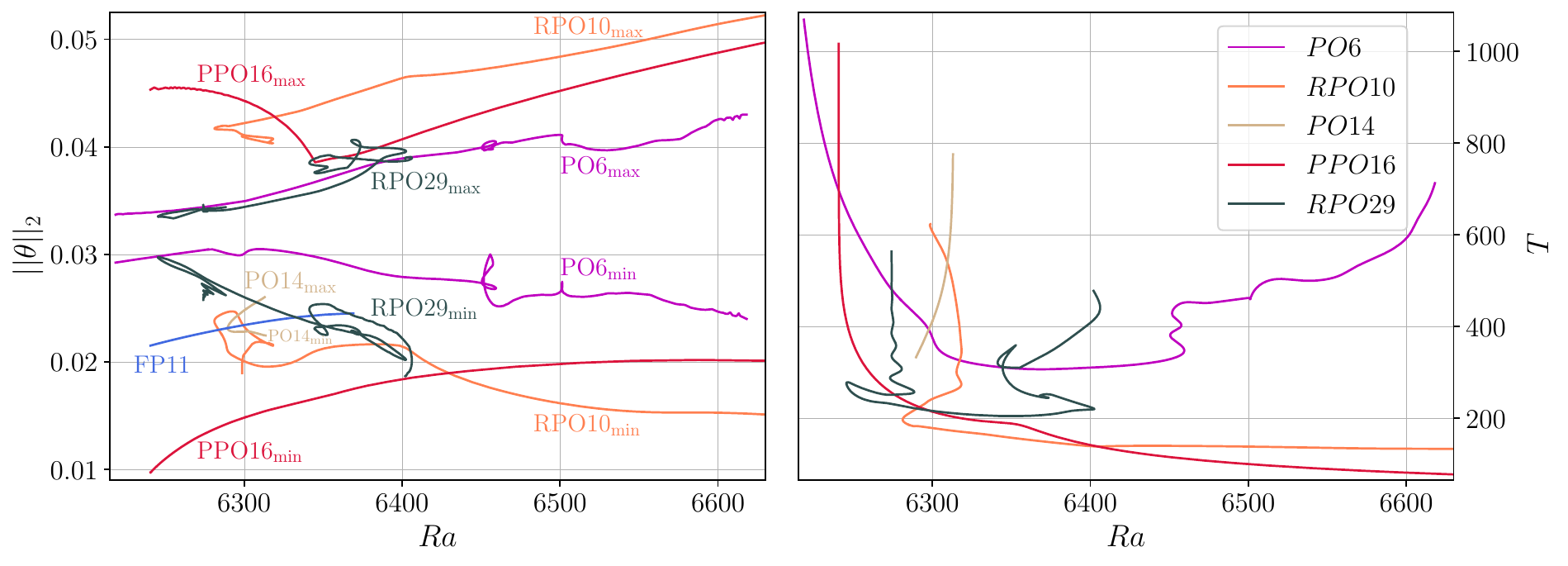}
    \captionsetup{font={footnotesize}}
    \captionsetup{width=13.5cm}
    \captionsetup{format=plain, justification=justified}
    \caption{\label{part3_sepa_BD_PO6_10_14_16_29} Temperature norms (left) and periods (right) of PO6, RPO10, PO14, PPO16 and RPO29. Branch PO6 approaches a heteroclinic cycle linking two symmetrically-related versions of FP2 in a global bifurcation at $Ra\approx6218.6$, at which its period diverges. At higher Rayleigh numbers, PO6 undergoes saddle--node bifurcations and continues to exist at least until $Ra=6615$. Branch RPO10 possibly bifurcates from FP4 in a global bifurcation at $Ra\approx 6298.7$ and continues to exist at least until $Ra=6650$. Branch PO14 bifurcates from FP11 in a Hopf bifurcation and terminates in a global bifurcation by meeting FP9. Branch PPO16 is created from FP9 in a global bifurcation at $Ra\approx6240.6$ and continues to exist until at least $Ra=6656.5$. Branch RPO29 bifurcates from FP4 at $Ra \approx 6274.14$ and terminates on FP2 at $Ra\approx 6402$ in two global bifurcations.}
\end{figure}

\begin{figure}
    \centering
    \includegraphics[width=\columnwidth]{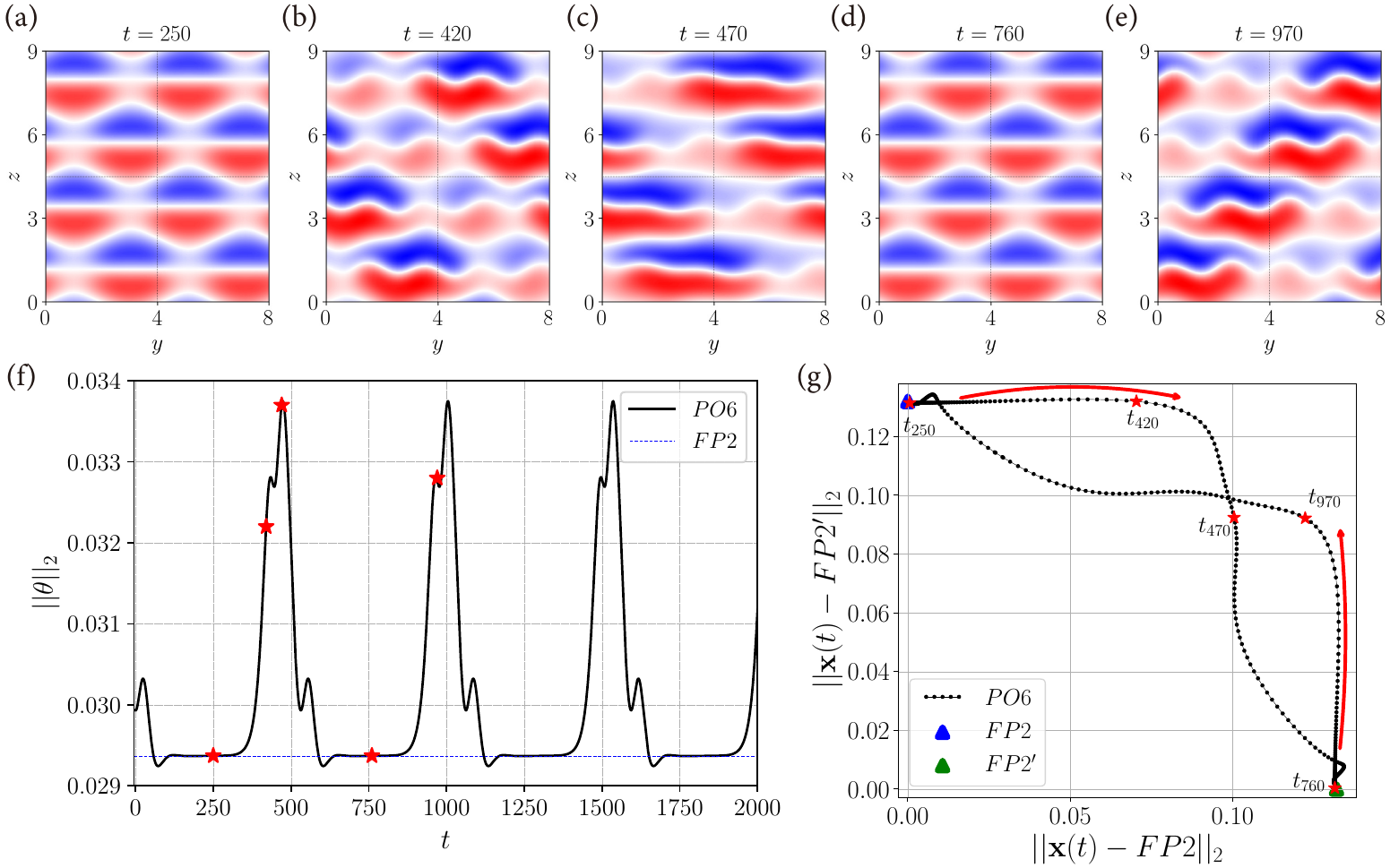}
    \captionsetup{font={footnotesize}}
    \captionsetup{width=13.5cm}
    \captionsetup{format=plain, justification=justified}
    \caption{\label{part3_PO6hetero} (a--e) Snapshots of the dynamics of PO6 at $Ra=6218.6$. Snapshots (a) and (d) show states which are close to two symmetry-related versions of FP2. (f) Time series of PO6 at $Ra=6218.6$ (with period $T= 1069.1$). (g) Phase space projection at $Ra=6218.6$ close to the global bifurcation point. The curve shows PO6 and triangles show two symmetry-related FP2 states involved in the heteroclinic cycle. In (f) and (g), the five red stars indicate the moments at which the snapshots (a)--(e) are taken. In (g), the red arrows show the direction of the trajectory.}
\end{figure}

\par We first observed PO6 at $Ra=6280$. Orbit PO6 has the spatio-temporal symmetry
\begin{equation}
(u,v,w,\theta)(x,y,z,t+T/2) = \pi_y(u,v,w,\theta)(x,y+L_y/2,z,t),
\end{equation}
should be understood as a pre-periodic orbit and be properly called PPO6. Exceptionally, we use PO6 here because we have converged this orbit with its full period and for consistency with PO2 in \citet{Zheng2024part2} whose dynamics closely resembles PO6. The period of PO6 shown in figures \ref{part3_BD-newPO}(b) and \ref{part3_sepa_BD_PO6_10_14_16_29} is twice its pre-period. Continuing PO6 backwards, it approaches a heteroclinic cycle linking two symmetrically-related versions of FP2. From $Ra=6280$ to $Ra=6218.6$, the period of PO6 increases monotonically to $T=1069.1$. (We have omitted the range $700<T<1069.1$ in figure \ref{part3_BD-newPO}(b) to better show the periods of other orbits.) At the global bifurcation point at slightly lower Rayleigh number, the period should diverge. Compare with, for example, the slope of PO2, PO14, PPO16, RPO18 and RPO29. We have been unable to continue PO6 further due to limits on numerical precision.

\par
\begin{figure}
    \centering
    \includegraphics[width=\columnwidth]{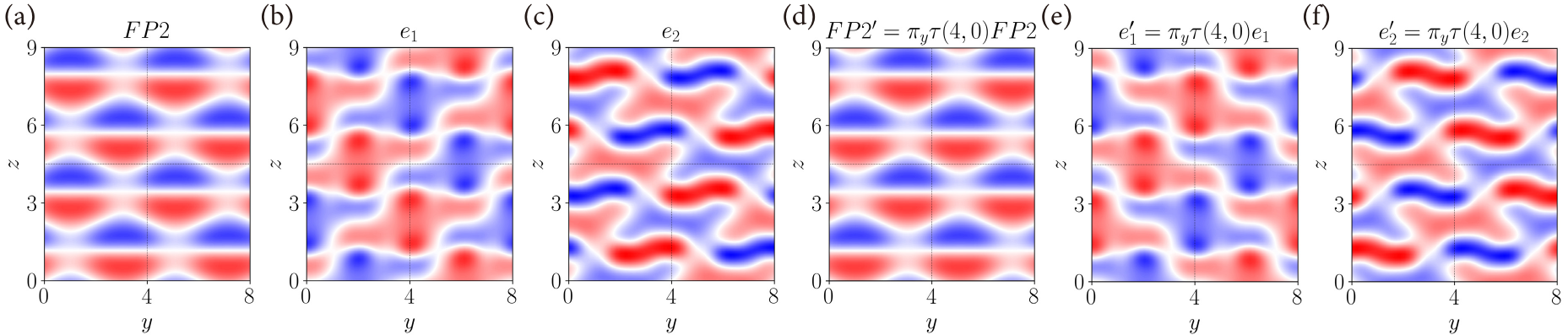}
    \captionsetup{font={footnotesize}}
    \captionsetup{width=13.5cm}
    \captionsetup{format=plain, justification=justified}
    \caption{\label{part3_PO6robustness} Equilibria and eigenmodes at $Ra=6218.6$. (a) FP2, (b) its unstable eigenmode $e_1$ and (c) its stable eigenmode $e_2$. (d) FP2$^\prime\equiv\pi_y\tau(4,0)$FP2, (e) its unstable eigenmode  $e_1^\prime\equiv\pi_y\tau(4,0)e_1$, and (f) its stable eigenmode $e_2^\prime\equiv\pi_y\tau(4,0)e_2$. The wavenumbers of the equilibria and eigenmodes in the $y$ direction suggest a 1:2 mode interaction.}
\end{figure}

 The dynamics of PO6 at $Ra=6218.6$, close to the global bifurcation point, is shown in figures \ref{part3_PO6hetero}(a-e). Orbit PO6 resembles RPO18 above. The dynamics of each half period of PO6 has four-fold translation symmetry along one of the diagonals, either $\braket{\tau(L_y/4, L_z/4)}$ or $\braket{\tau(L_y/4, -L_z/4)}$; compare figures \ref{part3_PO6hetero}(b) and \ref{part3_PO6hetero}(e). The only symmetry possessed instantaneously by all members of PO6 is $\braket{\tau(L_y/2, L_z/2)}$. The heteroclinic bifurcation is evidenced by the plateaus in the time series in figure \ref{part3_PO6hetero}(f) as well as by the phase space projection in figure \ref{part3_PO6hetero}(g). The eigenvectors along which PO6 approaches and escapes from FP2 (and FP2$^\prime\equiv\pi_y\tau(L_y/2,0)$FP2) are shown in figure \ref{part3_PO6robustness}, together with the phase of FP2 (and FP2$^\prime$).

\par To show that PO6 approaches a robust heteroclinic cycle as is the case for PO2, we identify two subspaces within the symmetry space of FP2:
\begin{equation}
\begin{array}{ll}
 S &\equiv {\rm Fix}|_{\braket{\pi_y\pi_{xz}, \tau(L_y/4,-L_z/4)}},\\
 S^\prime &\equiv {\rm Fix}|_{\braket{\pi_y\pi_{xz}, \tau(L_y/4,L_z/4)}}.
\end{array}
\end{equation}
For the flow restricted to subspace $S$ ($S^\prime$), FP2 is a saddle (sink), FP2$^\prime$ is a sink (saddle), and there exists a saddle-sink connection FP2$\rightarrow$FP2$^\prime$ (FP2$^\prime$$\rightarrow$FP2). More details of the conditions required for a robust heteroclinic cycle can be found in \citet{Krupa1997, Reetz2020a} and \S 4.2.3 of \citet{Zheng2024part2}. Like PO2, this robust cycle results from a 1:2 mode interaction \citep{Armbruster1988}. As the arguments on robustness and 1:2 interaction are fundamentally the same as for PO2, in order to avoid repetition, we refer readers to \S 4.2.3 of our previous paper \citep{Zheng2024part2}.

\par We have continued PO6 forward in Rayleigh number up to $Ra=6618.12$ where it has period $T=712.76$. Along the branch, PO6 undergoes a sequence of saddle--node bifurcations between $6450\lesssim Ra \lesssim6460$ and then at $Ra\approx6505$, as shown in figure \ref{part3_sepa_BD_PO6_10_14_16_29}. Beyond $Ra=6550$, the period of PO6 increases monotonically and seems to diverge. Integrating PO6 at $Ra=6618.12$ forward in time, we observe that its trajectory tends to visit two symmetrically-related instances of FP2, with the time spent near FP2 substantially shorter than at $Ra= 6218.6$. We suggest that PO6 may bifurcate again from FP2 in another heteroclinic bifurcation slightly beyond $Ra=6618.12$. However, since the continuation becomes computationally difficult for higher Rayleigh numbers, we have not been able to confirm this. We do not show snapshots of PO6 at $Ra=6618.12$, as they do not differ substantially from those in figures \ref{part3_PO6hetero}(a-e).

\subsubsection{Orbit PPO7: saddle--node bifurcations and isola}
\label{part3_UPO7}
\begin{figure}
    \centering
    \includegraphics[width=\columnwidth]{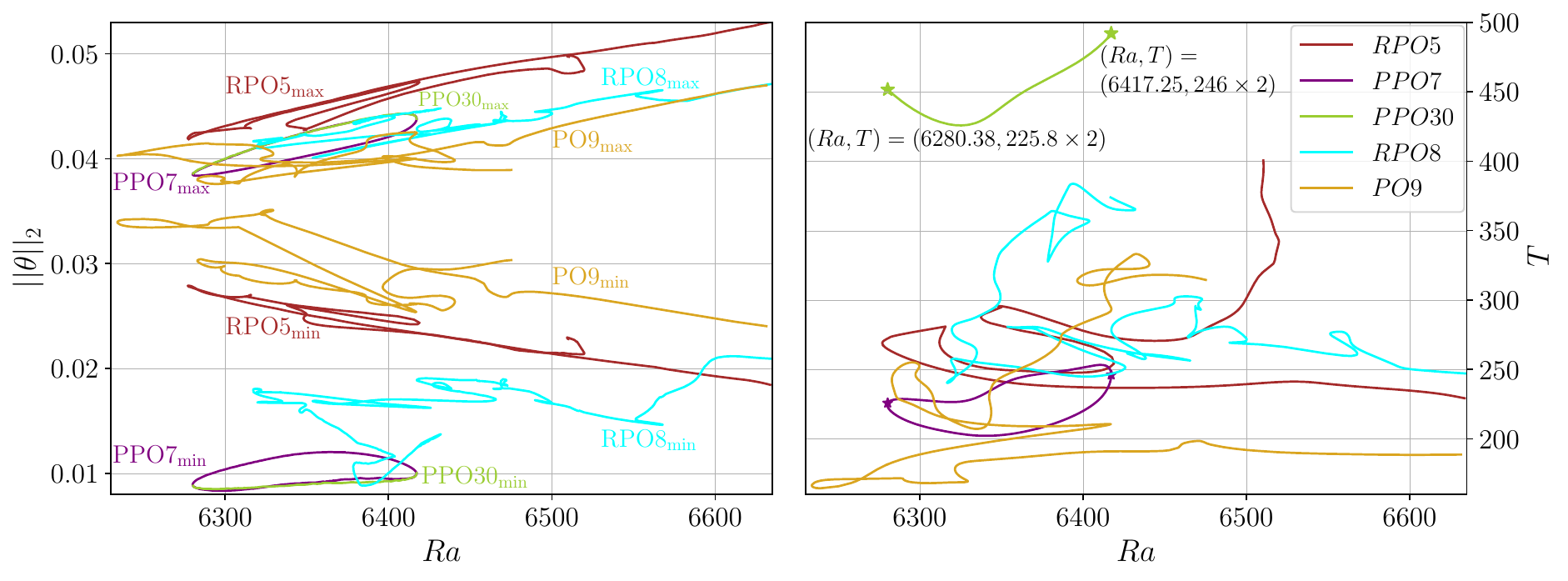}
    \captionsetup{font={footnotesize}}
    \captionsetup{width=13.5cm}
    \captionsetup{format=plain, justification=justified}
    \caption{\label{part3_sepa_BD_PO5_7_8_9_30} Temperature norms (left) and periods (right) of RPO5, PPO7, RPO8, PO9 and PPO30. Branches RPO5, RPO8 and PO9 undergo saddle--node bifurcations and are continued until $Ra=6635$ for one of their endpoints. At the other endpoints, RPO5 possibly bifurcates from FP4 in a global bifurcation, and the termination of RPO8 and PO9 are unclear. Branch PPO7 undergoes saddle--node bifurcations and forms an isola. Branch PPO30 bifurcates from and terminates on PPO7 in two period-doubling bifurcations.}
\end{figure}

\begin{figure}
    \centering
    \includegraphics[width=\columnwidth]{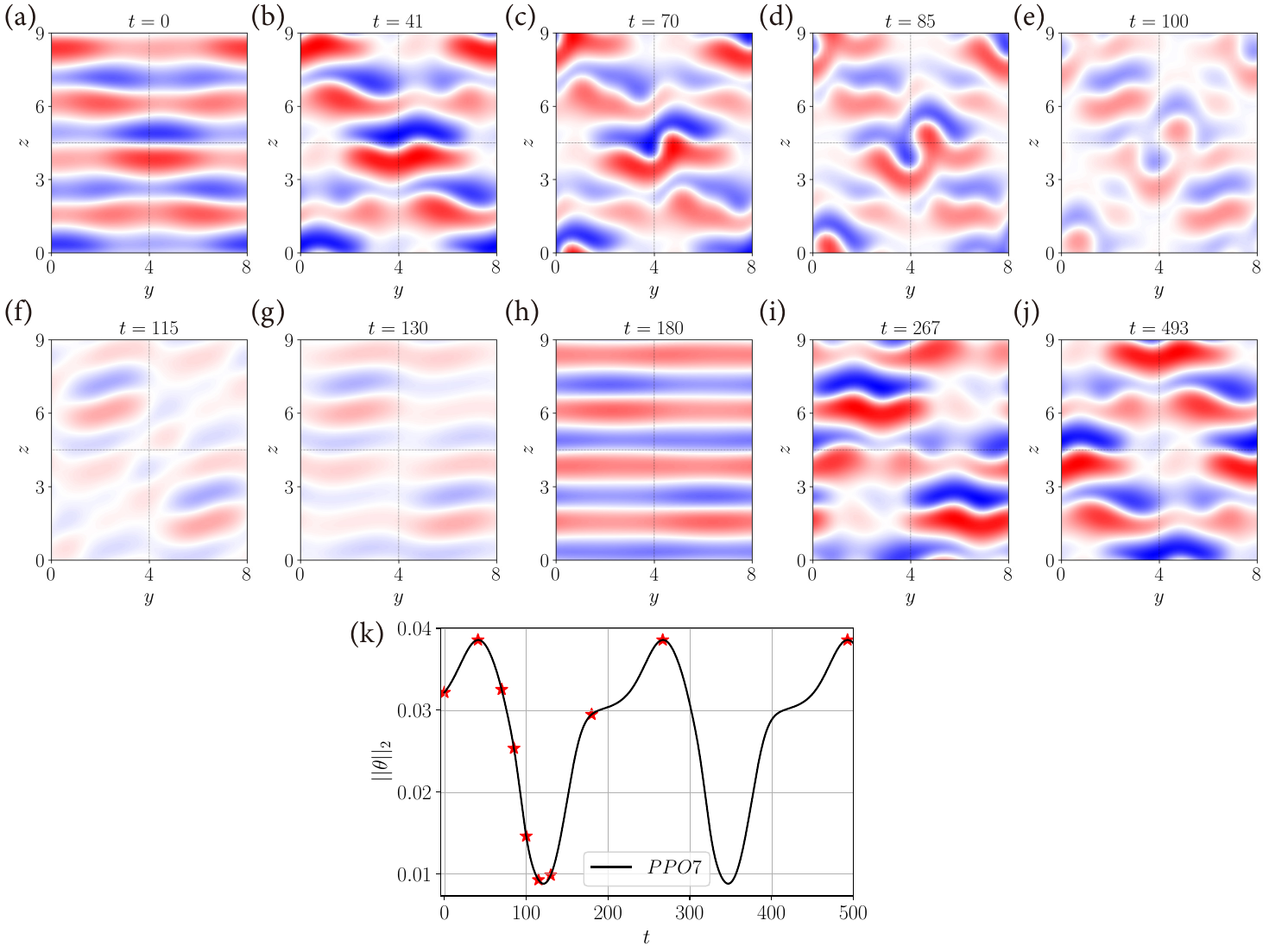}
    \captionsetup{font={footnotesize}}
    \captionsetup{width=13.5cm}
    \captionsetup{format=plain, justification=justified}
    \caption{\label{part3_PO7_series_snapshots} Dynamics of PPO7 at $Ra=6280.38$ with pre-period $T=226$. (a--j) Snapshots of the midplane temperature field. (k) Time series from DNS. The ten red stars indicate the moments at which the snapshots (a)--(j) are taken.}
\end{figure}

\par Like RPO15, RPO17, RPO26 and RPO28 that are discussed in \S \ref{part3_isola_RPO15-17-26-28}, branch PPO7 also forms an isola, with two saddle--node bifurcations at $Ra=6280.1$ and $6417.2$, as shown in figure \ref{part3_sepa_BD_PO5_7_8_9_30}. Orbit PPO7 gives rise to PPO30 via two period-doubling bifurcations that will be discussed later in \S \ref{part3_UPO30}.

\par The dynamics of PPO7 at $Ra=6280.38$ (the period-doubling bifurcation point) is shown in figures \ref{part3_PO7_series_snapshots}(a-j). The simulation starts from a state close to the moustache rolls (FP5 in \citet{Zheng2024part2}); the roll at the domain centre (as well as corners due to $\tau(L_y/2, L_z/2)$ symmetry) then becomes more intense, distorted and ramified (figures \ref{part3_PO7_series_snapshots}b-e). At $t=115$, the central roll pinches off and merges with neighbouring rolls at $t=130$. After a smooth transition towards nearly straight rolls at $t=180$, the trajectory returns to the distorted-roll state at $t=267$. 

\par Orbit PPO7 has the spatio-temporal symmetry:
\begin{align}
(u,v,w,\theta)(x,y,z,t+T) &= \pi_y(u,v,w,\theta)(x, y+L_y/4, z-L_z/4, t),
\end{align}
where $T=226$ is the pre-period of PPO7 at $Ra=6280.38$. After a pre-period, the state at $t=267$, figure \ref{part3_PO7_series_snapshots}(i) is a reflected and translated version of the state at $t=41$, figure \ref{part3_PO7_series_snapshots}(b); after integrating over a second pre-period, the states at $t=493$ and $t=41$ are related by $\sigma \equiv \tau(\pm L_y/2,0)$ or $\tau(0, \pm L_z/2)$. Finally, after integrating during four pre-periods, the initial state matches the final state, i.e.\ PPO7 is a periodic orbit. 

\par A remarkable feature of PPO7 in figure \ref{part3_BD-newPO}(a) is that its minimum temperature norm along the branch ($\lvert\lvert \theta \lvert\lvert_2 \approx 0.01$) is almost the lowest among all orbits. The faint figures \ref{part3_PO7_series_snapshots}(f-g) correspond to the moments of lowest temperature norm. These are the moments of cutting-joining-like dynamics of convection rolls, very similar to the longitudinal burst pattern observed by \citet{Daniels2000} experimentally and by \citet{Reetz2020b} numerically, at slightly different control parameters. The cutting-joining dynamics induces defects, disordered structures in rolls and roll-bursting; these contribute to the transition to turbulence. To the best of our knowledge, PPO7 may be the first (pre-)periodic orbit that captures these aspects of the dynamics.

\subsubsection{Orbit RPO10: saddle--node and global bifurcations}
\label{part3_UPO10}
\begin{figure}
    \centering
    \includegraphics[width=\columnwidth]{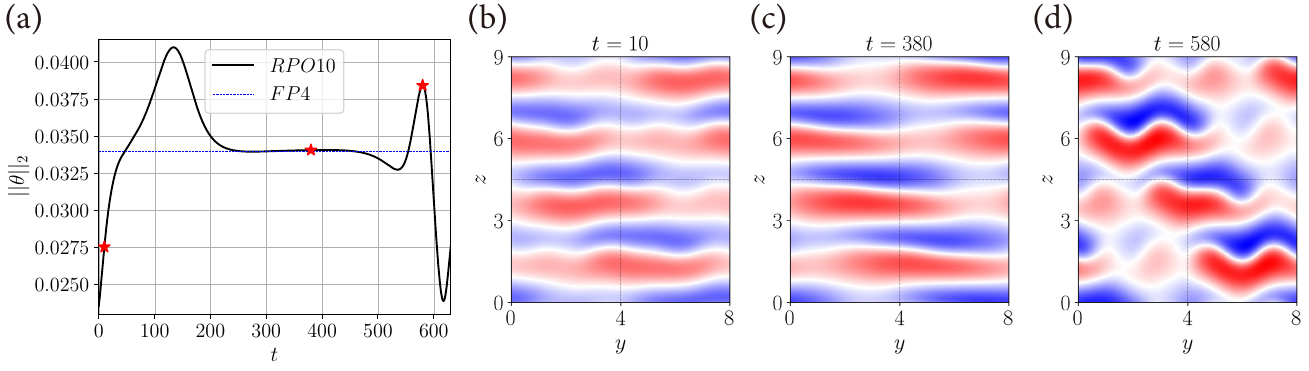}
    \captionsetup{font={footnotesize}}
    \captionsetup{width=13.5cm}
    \captionsetup{format=plain, justification=justified}
    \caption{\label{part3_PO10_series_snapshots} Dynamics of RPO10 at $Ra=6298.686$ with relative period $T=623.35$. (a) Time series of RPO10; the three red stars indicate the moments at which the snapshots (b)--(d) of the midplane temperature field are taken.}
\end{figure}

\par We first found RPO10 at $Ra=6400$. Forward continuation in $Ra$ shows that it exists until at least $Ra=6675$ where we stopped the continuation. Continuing RPO10 backwards in $Ra$, it undergoes a sequence of saddle--node bifurcations after which its period increases monotonically, as evidenced by figure \ref{part3_sepa_BD_PO6_10_14_16_29}. We have been able to continue RPO10 until $Ra=6298.686$ with relative period $T=623.35$, shortly after the last saddle--node bifurcation at $(Ra,T)=(6298.39,622.97)$. 

\par Integrating RPO10 at $Ra=6298.686$ in time, we observe a long plateau ($250<t<460$) in the time series shown in figure \ref{part3_PO10_series_snapshots}(a). The dynamics of RPO10 at this $Ra$ is shown in figures \ref{part3_PO10_series_snapshots}(b-d). The states corresponding to the location of the plateau are very similar to FP4, in terms of both flow structure and temperature norm. This suggests that RPO10 disappears in a global homoclinic bifurcation by meeting FP4, although the slope of the last computed portion of $T(Ra)$ in figure \ref{part3_sepa_BD_PO6_10_14_16_29} is not as close to vertical as the corresponding slopes for PO6, PO14, PPO16 and RPO29.

\subsubsection{Orbit RPO25: pitchfork, saddle--node and period-halving bifurcations}
\label{part3_UPO25}
\par As shown in figure \ref{part3_sepa_BD_PO19_25}, RPO25 bifurcates from RPO19 at $Ra=6260.5$ in a pitchfork bifurcation, in which the translation symmetry $\tau(L_y/4, L_z/4)$ is broken to $\tau(L_y/2, L_z/2)$ and the reflection symmetry $\pi_y\pi_{xz}$ is retained. Orbit RPO25 then undergoes several saddle--node bifurcations and finally terminates in a period-halving bifurcation at $Ra=6432.34$ (indicated in figure \ref{part3_sepa_BD_PO19_25}). We do not show or discuss the period-halved orbit in this work.

\subsubsection{Orbit RPO27: pitchfork and saddle--node bifurcations}
\label{part3_UPO27}
\par As shown in figure \ref{part3_sepa_BD_PO17_18_27}, RPO27 bifurcates from RPO18 at $Ra=6279.7$ in a pitchfork bifurcation at which $\tau(L_y/4, L_z/4)$ symmetry is broken to $\tau(L_y/2, L_z/2)$ (with $\pi_y\pi_{xz}$ retained). It undergoes a sequence of saddle--node bifurcations, particularly between $Ra=6420$ and $6500$, and we have continued it until $Ra=6650$. 

\subsubsection{Orbit RPO29: saddle--node and global bifurcations}
\label{part3_UPO29}
\begin{figure}
    \centering
    \includegraphics[width=\columnwidth]{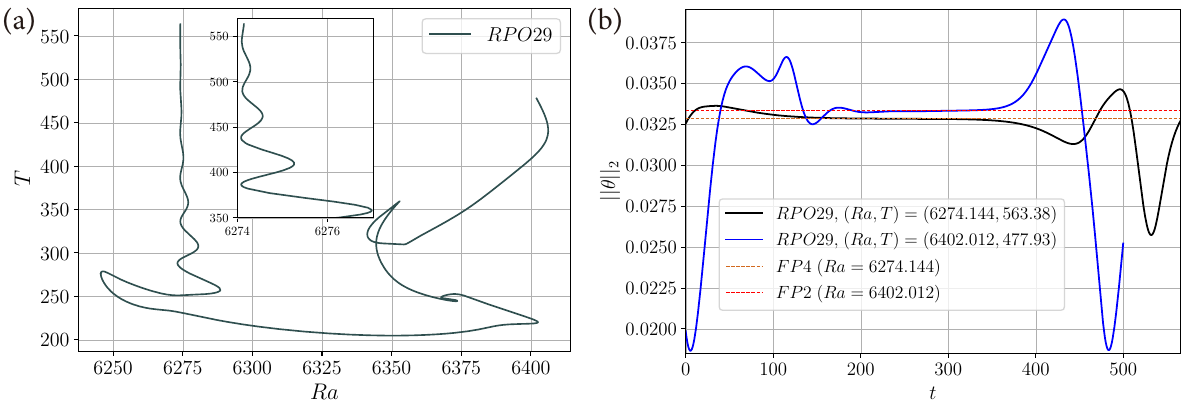}
    \captionsetup{font={footnotesize}}
    \captionsetup{width=13.5cm}
    \captionsetup{format=plain, justification=justified}
    \caption{\label{part3_PO29_series_period} (a) Periods and (b) time series of RPO29. (Branch RPO29 also appears as part of figure \ref{part3_sepa_BD_PO6_10_14_16_29}.) The inset in (a) shows a sequence of saddle--node bifurcations before the global bifurcation at $Ra\approx6274.14$. (b) Time series from the last continuation point (longest period) at $Ra=6274.144$ and $Ra=6402.012$. Branch RPO29 approaches FP2 and FP4 in two different global homoclinic bifurcations at its two endpoints.}
\end{figure}

\par We first observed RPO29 at $Ra=6300$. Its bifurcation diagram is shown in figure \ref{part3_sepa_BD_PO6_10_14_16_29}; we show its period again in figure \ref{part3_PO29_series_period}(a). Backward continuation reveals the interesting feature of at least 13 saddle--node bifurcations (and probably even more if it could be continued further towards higher periods) before the global bifurcation at which the period diverges. We have been able to continue RPO29 down to $Ra=6274.144$, where it has relative period $T=563.38$. Orbit RPO29 undergoes successive approaches to FP4 (see time series in figure \ref{part3_PO29_series_period}b), as is the case for RPO10 in \S \ref{part3_UPO10}.

\par Multiple regularly spaced saddle--node bifurcations, as in figure \ref{part3_PO29_series_period}(a), are seen in at least two situations. One the homoclinic snaking of branches of spatially localized equilibria \citep{burke2006localized, batiste2006spatially} and periodic orbits \citep{AlSaadi2024}. At each saddle--node bifurcation, a new pair of rolls or structures is added to the solution. For an increase in periods, as in figure \ref{part3_PO29_series_period}(a), we could conjecture an increasing number of temporal maxima. However, in our case, we have checked that the states along the RPO29 branch do not show an increasing number of rolls or structures, nor an increasing number of peaks after each saddle--node bifurcation. 

\par Another situation in which multiple regularly spaced saddle--node bifurcations of periodic orbits occur is the Shil'nikov bifurcation. This is the approach to a homoclinic orbit that is formed at the collision of a limit cycle with a saddle--focus which has one real eigenvalue and one complex eigenpair \citep{Glendinning1984}. In our case, RPO29 and FP4 are in the symmetry subspace $\braket{\pi_{y}\pi_{xz}, \tau(L_y/2, L_z/2)}$ and hence we consider the leading eigenvalues of FP4 in this symmetry subspace. At $Ra=6274.144$, these are $[\lambda_1, \lambda_2, \lambda_3, \lambda_4, \lambda_{5,6}] = [0.0384, 0.0364, 0.0096, 0.0019, -0.0172 \pm 0.1167i]$. We have determined that the eigenvalues controlling the escape from and approach to FP4 are those closest to zero, $\lambda_4$ and $\lambda_{5,6}$. \cite{Glendinning1984} show that multiple saddle--node bifurcations occur if the rate of escape exceeds the rate of approach, but in our case, the rate of approach exceeds the rate of escape by a factor of $|-0.0172/0.0019| \approx 9$. This implies that the oscillations in figure \ref{part3_PO29_series_period}(a) will cease when the period becomes large enough, to be replaced by a monotonic increase in period. Unfortunately, we have not been able to continue RPO29 further towards higher periods and thus cannot confirm the potential monotonic increase.

\par Continuing RPO29 forward from $Ra=6300$, many saddle--node bifurcations are also seen and the branch also terminates in a global homoclinic bifurcation, this time by meeting FP2. We have been able to continue RPO29 until $(Ra,T)=(6402.012, 477.93)$. Even though we believe that this is still far from the actual global bifurcation point as the period is not yet very long, a close approach to FP2 is evidenced by a clear plateau (whose corresponding norm is very close to that of FP2) in the time series in figure \ref{part3_PO29_series_period}(b) and by inspection of flow fields (not shown).

\subsubsection{Orbit PPO30: period-doubling bifurcations}
\label{part3_UPO30}
\par Orbit PPO30 bifurcates from and terminates on PPO7 in two period-doubling bifurcations, at $Ra=6280.38$ and $6417.25$, indicated in figure \ref{part3_sepa_BD_PO5_7_8_9_30}. For PPO30,
\begin{align}
    (u,v,w,\theta)(x,y,z,t+T) &= (u,v,w,\theta)(x,y\pm L_y/2,z,t), \nonumber \\
    &=(u,v,w,\theta)(x,y,z \pm L_z/2,t),
\end{align}
where $T$ is the pre-period of PPO7. After two pre-periods, the initial state matches the final one. The quantities $\lvert\lvert \theta \lvert\lvert_2$ of PPO7 and PPO30 are almost indistinguishable as can be seen in figure \ref{part3_sepa_BD_PO5_7_8_9_30} and as is usual for period-doubling bifurcations.

\subsection{Symmetry subspace: four-fold translation}
\label{part3_sym_4fold}
\par In this subsection, we discuss two time-periodic solutions (RPO12 and RPO20) identified in the symmetry subspace $\braket{\tau(L_y/4, L_z/4)} \sim Z_4$. Branches RPO12 and RPO20 are shown in the bifurcation diagram of figure \ref{part3_sepa_BD_PO11_12_20}.

\subsubsection{Orbit RPO12: saddle--node and global bifurcations}
\label{part3_UPO12}
\begin{figure}
    \centering
    \includegraphics[width=\columnwidth]{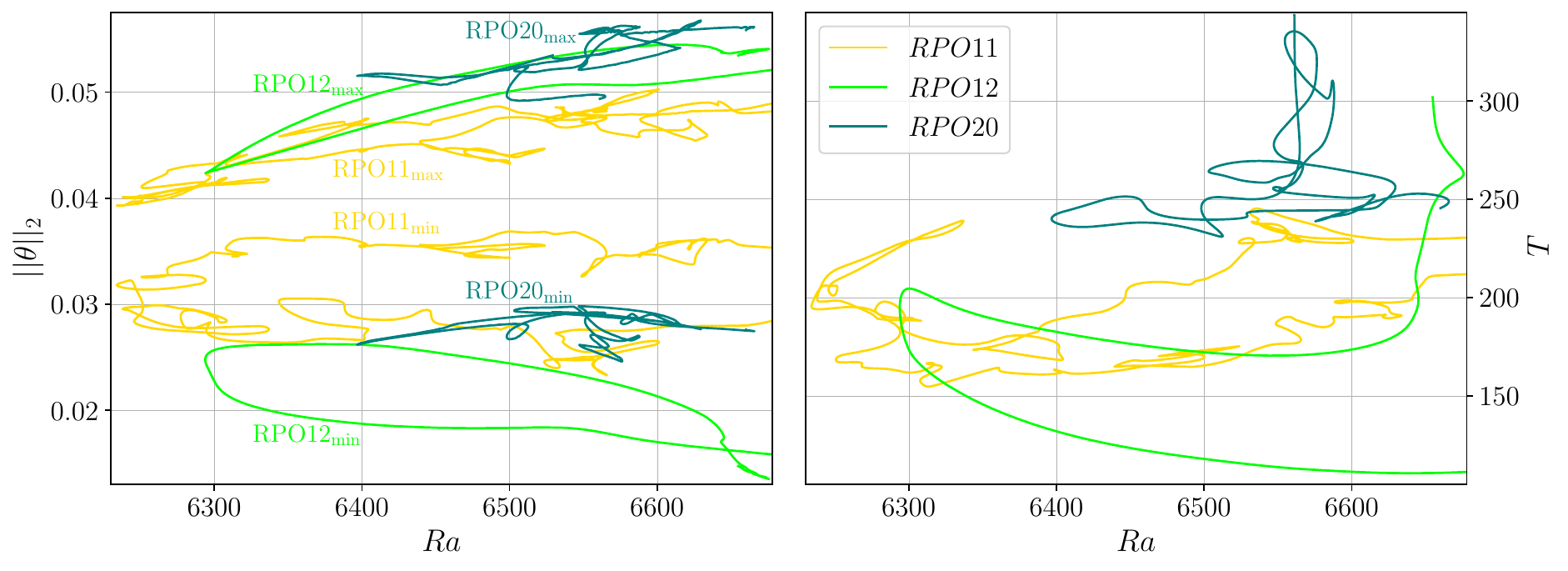}
    \captionsetup{font={footnotesize}}
    \captionsetup{width=13.5cm}
    \captionsetup{format=plain, justification=justified}
    \caption{\label{part3_sepa_BD_PO11_12_20} Temperature norms (left) and period (right) of RPO11, RPO12 and RPO20. Branch RPO11 undergoes saddle--node bifurcations and both the lower and upper branches are continued beyond $Ra=6680$; its bifurcation structure remains unclear. The lower RPO12 branch exists beyond $Ra=6680$, while the upper branch seems to terminate in a global bifurcation by meeting FP4, close to $Ra=6655$. Branch RPO20 bifurcates from FP4 in a global bifurcation at $Ra\approx6561$ at which its period seems to diverge; its termination is unclear.}
\end{figure}

\begin{figure}
    \centering
    \includegraphics[width=\columnwidth]{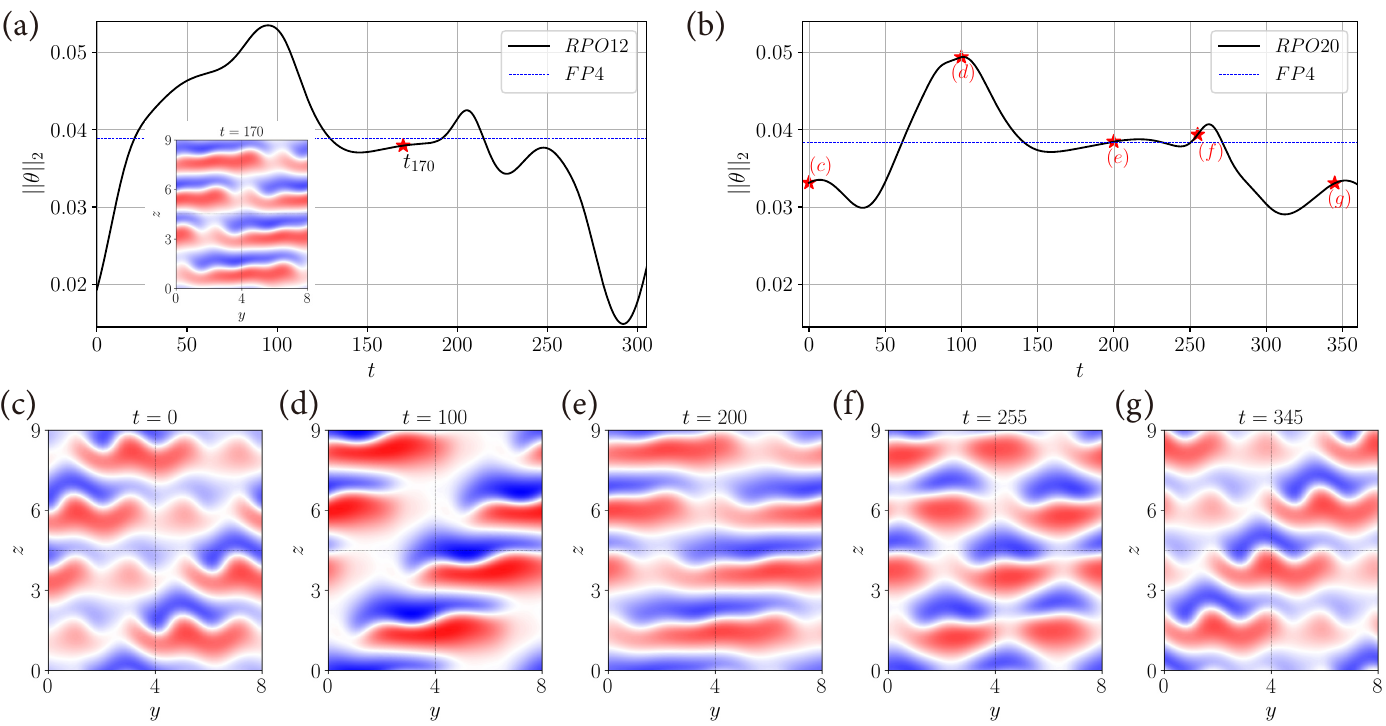}
    \captionsetup{font={footnotesize}}
    \captionsetup{width=13.5cm}
    \captionsetup{format=plain, justification=justified}
    \caption{\label{part3_PO20_series_snapshots} (a) Time series of RPO12 with relative period $T=301.9$ at $Ra=6654.865$. A snapshot of the midplane temperature field at instant $t=170$ is shown in the inset and is close to FP4. (b) Time series of RPO20 with relative period $T=343$ at $Ra=6561.2$. The five red stars indicate the moments at which the snapshots (c)--(g) are taken. Snapshot (e) is close to FP4.}
\end{figure}

\par Orbit RPO12 undergoes saddle--node bifurcations at $Ra=6293.9$, $6643 \lesssim Ra \lesssim 6646$ and at $Ra=6675.8$. The lower branch (in terms of period) of RPO12 has been continued until $(Ra, T)=(6680,111.4)$ where we stopped the continuation. The upper branch is continued until $Ra=6654.865$ where the relative period $T=301.9$ is the highest that we were able to attain numerically. 

\par Integrating RPO12 at $Ra=6654.865$, we observe that the dynamics slows down slightly near a state that is close to FP4; see figure \ref{part3_PO20_series_snapshots}(a) and its inset. We subsequently used this state ($t=170$) as the initial guess for Newton's method to converge to FP4 at $Ra=6654.865$. Even though the plateau in figure \ref{part3_PO20_series_snapshots}(a) is not obvious and the period of RPO12 in figure \ref{part3_sepa_BD_PO11_12_20} is not yet very long, our observations suggest that RPO12 terminates in a global homoclinic bifurcation by meeting FP4. This scenario is very similar to that of RPO20 to be discussed just after in \S \ref{part3_UPO20}, for which we will show more snapshots.

\subsubsection{Orbit RPO20: saddle--node and global bifurcations}
\label{part3_UPO20}
\par We have continued RPO20 to $Ra=6561.2$ with relative period $T=343.6$ and to $Ra=6660.2$ with $T=245.5$. Close to $Ra=6561.2$, its period seems to diverge, slightly more so than that of RPO12. Figure \ref{part3_PO20_series_snapshots}(b) shows the time series from a simulation of RPO20 at $Ra=6561.2$, with the corresponding snapshots of the temperature field shown in figures \ref{part3_PO20_series_snapshots}(c-g). The plateau-like behaviour ($150 \lesssim t \lesssim 250$) in the time series as well as the close resemblance between figure \ref{part3_PO20_series_snapshots}(e) and FP4 suggest that RPO20 bifurcates from FP4 in a global homoclinic bifurcation at a nearby Rayleigh number. 

\par By looking at the snapshots in figures \ref{part3_PO20_series_snapshots}(c-g), we notice that the dynamics is not very different from previous cases, as the reflection symmetry $\braket{\pi_y\pi_{xz}}$ is only weakly broken by the global bifurcation from FP4. Since it is clear that the RPO20 (and RPO12 in \S \ref{part3_UPO12}) with the longest period that we have succeeded in computing is still far from the actual homoclinic cycle, we do not discuss or show the eigendirections of FP4 along which RPO20 (and RPO12) may approach and escape from. The termination of the other end of the RPO20 branch is unclear and not discussed.

\subsection{Symmetry subspace: two-fold translation}
\label{part3_sym_2fold}
\par
\begin{figure}
    \centering
    \includegraphics[width=\columnwidth]{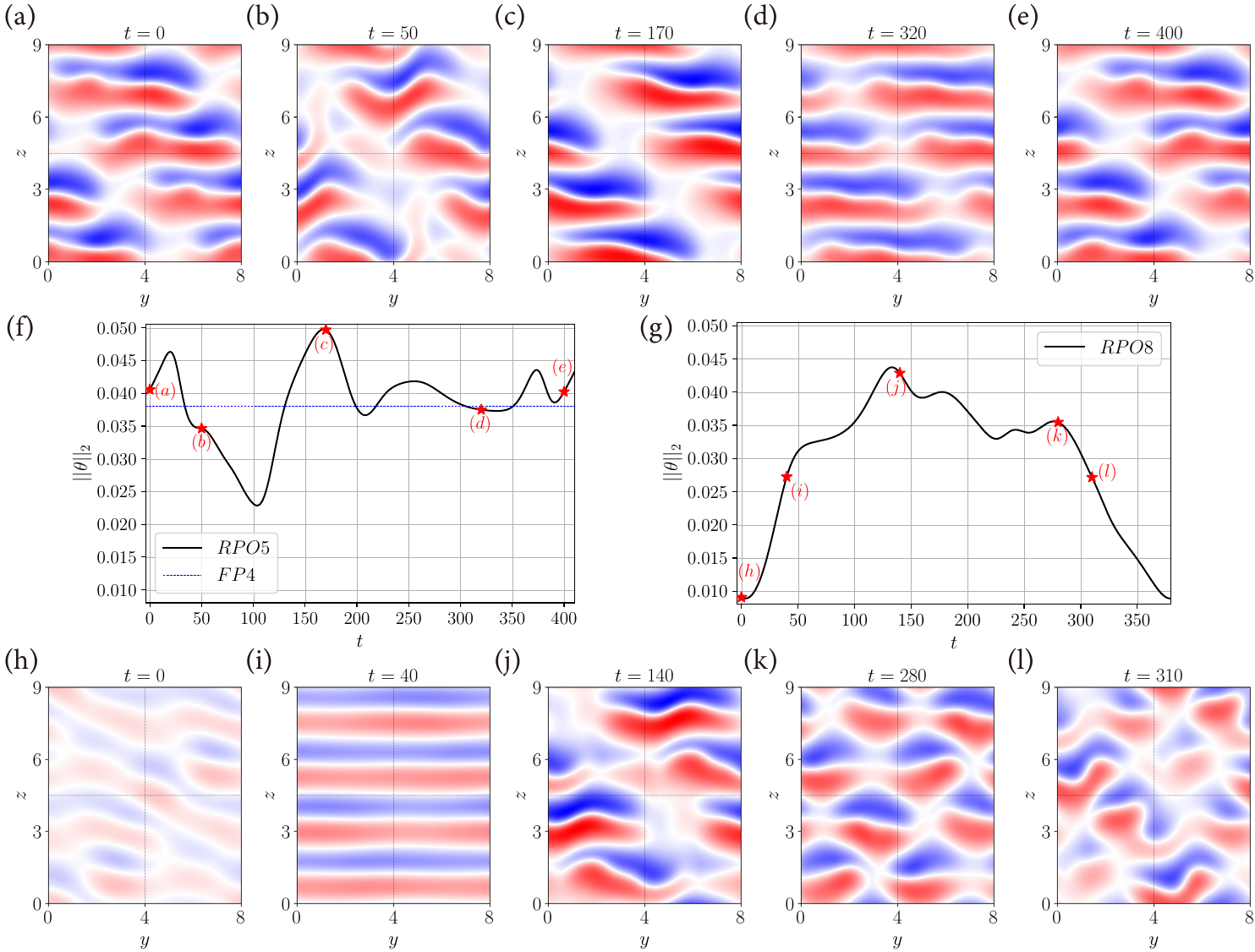}
    \captionsetup{font={footnotesize}}
    \captionsetup{width=13.5cm}
    \captionsetup{format=plain, justification=justified}
    \caption{\label{part3_PO5PO8_series_snapshots} Dynamics of RPO5 with relative period $T=400.5$ at $Ra=6510.4$ and of RPO8 with $T=375.4$ at $Ra=6388.46$. (a--e, h--l) Snapshots of the midplane temperature field. Snapshot (d) of RPO5 is similar to FP4 (figure \ref{part3_BD-newFP}d) and converges to FP4 when used as an initial guess for Newton's method. (f--g) Time series, initialized by the states shown in (a) and (h).}
\end{figure}

 In this subsection, we discuss two relative periodic orbits (RPO5 and RPO8) identified in the symmetry subspace $\braket{\tau(L_y/2, L_z/2)} \sim Z_2$. Their branches are contained in the bifurcation diagram of figure \ref{part3_sepa_BD_PO5_7_8_9_30}, while their time series and snapshots are shown in figure \ref{part3_PO5PO8_series_snapshots}.

\subsubsection{Orbit RPO5: saddle--node and global bifurcations}
\label{part3_UPO5}
\par In the $Ra$ range we study, RPO5 undergoes a sequence of saddle--node bifurcations. The lower branch (in period) continues to exist until at least $Ra=6635$. For the upper branch, the seemingly diverging period at $Ra\approx6510.4$ suggests that RPO5 might disappear in a global bifurcation. We have been able to continue RPO5 until $Ra=6510.4$ with relative period $T=400.5$. Integrating RPO5 at $Ra=6510.4$ in time, the dynamics slows down slightly close to FP4 ($t\approx300$), as shown by the time series in figure \ref{part3_PO5PO8_series_snapshots}(f) and the snapshot in figure \ref{part3_PO5PO8_series_snapshots}(d). We expect that the time spent near FP4 would increase if we were able to continue RPO5 further.

\subsubsection{Orbit RPO8: saddle--node bifurcations}
\label{part3_UPO8}
\par As shown in figure \ref{part3_sepa_BD_PO5_7_8_9_30}, RPO8 undergoes a sequence of saddle--node bifurcations and is continued until $Ra=6636.26$ (lower branch, where relative period $T=246.97$) and $Ra=6416.88$ (upper branch, where $T=373.92$). With the available information, we have not been able to determine the origin of RPO8. Figures \ref{part3_PO5PO8_series_snapshots}(h-l) show five snapshots of RPO8 at $Ra=6388.46$. Like PPO7, rolls in RPO8 tend to distort and to develop defects, and the variation of $\lvert\lvert \theta \lvert\lvert_2$ along the orbit is large; compare for instance figures \ref{part3_PO5PO8_series_snapshots}(h) and (j). Note also that figure \ref{part3_PO5PO8_series_snapshots}(k) is similar to a less symmetric version of FP8 shown in figure \ref{part3_BD-newFP}(f).

\subsection{Symmetry subspace: reflection with three-fold translation}
\label{part3_sym_ref_3fold}
\par Only one orbit is identified in the symmetry subspace $\braket{\pi_{y}\pi_{xz}, \tau(L_y/3, L_z/3)} \sim D_3$.

\subsubsection{Orbit PO14: Hopf and global bifurcations}
\label{part3_UPO14}
\begin{figure}
    \centering
    \includegraphics[width=\columnwidth]{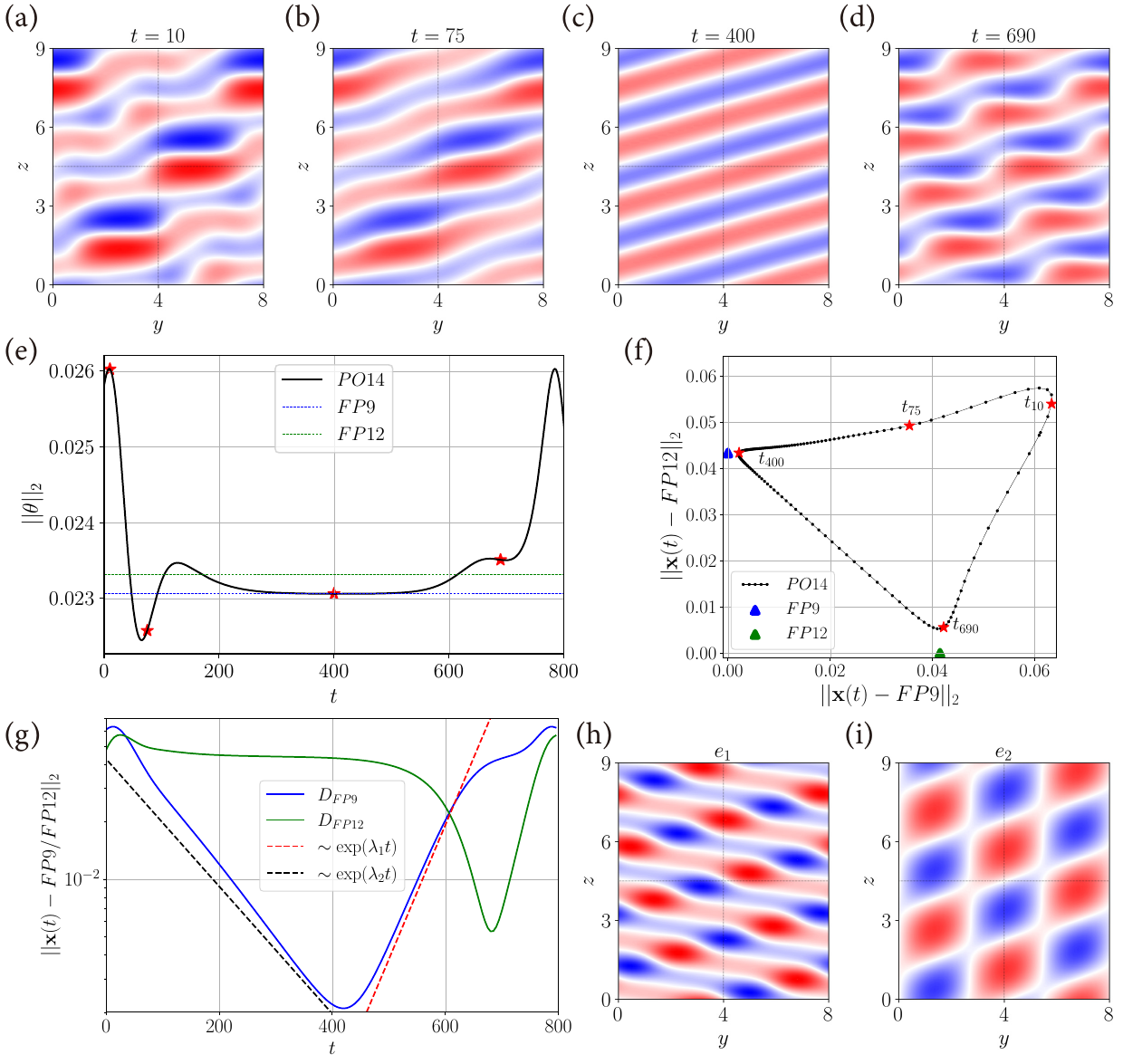}
    \captionsetup{font={footnotesize}}
    \captionsetup{width=13.5cm}
    \captionsetup{format=plain, justification=justified}
    \caption{\label{part3_PO14_series_snapshots} Dynamics of PO14 with period $T=775.62$ at $Ra=6313$ (close to the global bifurcation point). (a--d) Snapshots of the midplane temperature field. Snapshots (c) and (d) converge to FP9 and FP12 when used as initial guesses. (e) Time series from DNS. (f) Phase space projection: shown are PO14 (curve with dots) as well as FP9 and FP12 (triangles). In (e) and (f), the four red stars indicate the moments at which the snapshots (a)--(d) are taken. (g) $L_2$-distance between each instantaneous flow field of PO14 and FP9 (and FP12). The dynamics of PO14 is exponential for most of the cycle (blue curve). The approaching (black dashed line) and escaping (red dashed line) dynamics of PO14 with respect to FP9 are shown and are governed by two eigenvalues, $\lambda_1$ and $\lambda_2$, of FP9. (h--i) Two eigenmodes $e_1$ and $e_2$ of FP9, visualized via the midplane temperature field.}
\end{figure}  

\par As shown in figure \ref{part3_sepa_BD_PO6_10_14_16_29}, PO14 bifurcates from FP11 at $Ra=6289.6$ in a symmetry-preserving Hopf bifurcation. (For consistency with symmetry groups of other orbits, we do not introduce tilted coordinates for PO14 as we did for FP9--FP12 in \S \ref{part3_FP9_12}. It can be verified that FP9, FP11 and FP12 all have the symmetry $\braket{\pi_{y}\pi_{xz}, \tau(L_y/3, L_z/3)}$ in the $y$-$z$ coordinate.) The period of PO14 close to the bifurcation point closely matches the value predicted by the bifurcating imaginary eigenvalue pair of FP11. Equilibria FP9, FP10 and FP11 are called secondary, tertiary and quaternary states, respectively. Orbit PO14 can thus be called a quinary state. Forward continuation of PO14 suggests that it terminates in a global homoclinic bifurcation by meeting FP9, close to $Ra=6313$ at which its period diverges. 

\par Figures \ref{part3_PO14_series_snapshots}(a-d) show four snapshots of PO14 at $Ra=6313$, the highest Rayleigh number that we have reached. The corresponding time series and phase space projection are shown in figures \ref{part3_PO14_series_snapshots}(e-f), with special instants indicated by red stars. The long plateau ($250\lesssim t \lesssim 550$) in the time series and the clustering of points close to FP9 in the phase space projection suggest that PO14 approaches and slows down near FP9. Interestingly, the time series also shows another short plateau near $t\approx690$, whose corresponding state, shown in figure \ref{part3_PO14_series_snapshots}(d), resembles FP12 shown in figure \ref{part3_BD-newFP}(j). However, the non-negligible difference of norms in both figures \ref{part3_PO14_series_snapshots}(e-f) between FP12 and the state shown in figure \ref{part3_PO14_series_snapshots}(d) suggests that PO14 does not visit FP12 closely. It might be possible that, if we were able to continue PO14 further with longer periods, FP12 would be visited more closely by PO14. But based on the available data, we conclude that PO14 ends by approaching a homoclinic cycle by meeting FP9.

\par In order to analyse this homoclinic cycle, we have computed the spectrum of FP9 at $Ra=6313$. (Rather than referring back to figure \ref{part3_BD-newFP}(g), the reader can look at figure \ref{part3_PO14_series_snapshots}(c), which closely resembles FP9. For a detailed explanation of the correspondence between global bifurcations and the eigenvalues of the equilibria that are approached by the trajectories, see \cite{Zheng2024part2}.) Restricting the computation to the symmetry subspace $\braket{\pi_{y}\pi_{xz}, \tau(L_y/3, L_z/3)}$ gives three leading eigenvalues, all real: $[\lambda_1, \lambda_2, \lambda_3] = [0.0162, -0.0077, -0.012]$. Since we imposed FP9's symmetries in the Arnoldi iterations, the two neutral eigenvalues corresponding to translations in the periodic directions $y$ and $z$ are not present. We have determined that the eigendirections along which PO14 leaves and approaches FP9 are $e_1$ and $e_2$ associated with $\lambda_1$ and $\lambda_2$, respectively. These eigendirections are confirmed by subtracting FP9 from the instantaneous flow fields of PO14, and comparing the resulting fields with eigenmodes obtained by Arnoldi iterations, as well as by the close matches for the exponential growth and decay rate between PO14 and FP9, see figure \ref{part3_PO14_series_snapshots}(g). Since $\lambda_1>|\lambda_2|$, the homoclinic orbit is unstable, which is confirmed by the chaotic behaviour in the time series after time-integrating about two periods (not shown in figure \ref{part3_PO14_series_snapshots}e).

\par The two relevant eigendirections can be interpreted and analysed by comparing PO14 and FP9. Eigenmode $e_1$, shown in figure \ref{part3_PO14_series_snapshots}(h), breaks the $O(2)$ symmetry of FP9 along its straight and homogeneous rolls by introducing alternating red and blue patches in this tilted direction; these patches lead to wavy-roll structures. It is not difficult to imagine that superposing FP9 (figure \ref{part3_PO14_series_snapshots}c) and $e_1$ gives approximately figure \ref{part3_PO14_series_snapshots}(d). Eigenmode $e_2$, shown in figure \ref{part3_PO14_series_snapshots}(i), consists of a grid of red and blue rhombi, while figure \ref{part3_PO14_series_snapshots}(b) consists of rolls with bulges and constrictions. The colours of the rhombi are opposite to those of the bulges and the same as those of the constrictions. Thus, superposing $e_2$ on figure \ref{part3_PO14_series_snapshots}(b) reduces both distortions, restoring the broken symmetries of FP9.

\subsection{Symmetry subspace: reflection with five-fold translation}
\label{part3_sym_ref_5fold}
\par Three orbits identified in the symmetry subspace $\braket{\pi_{y}\pi_{xz}, \tau(L_y/5, L_z/5)} \sim D_5$ are discussed in this subsection. The dynamics of these three orbits appears to be similar and we only show snapshots of PPO16 (figures \ref{part3_PO16_series_snapshots}a-e) for illustration. 

\subsubsection{Orbit PPO16: global bifurcation}
\label{part3_UPO16}
\begin{figure}
    \centering
    \includegraphics[width=\columnwidth]{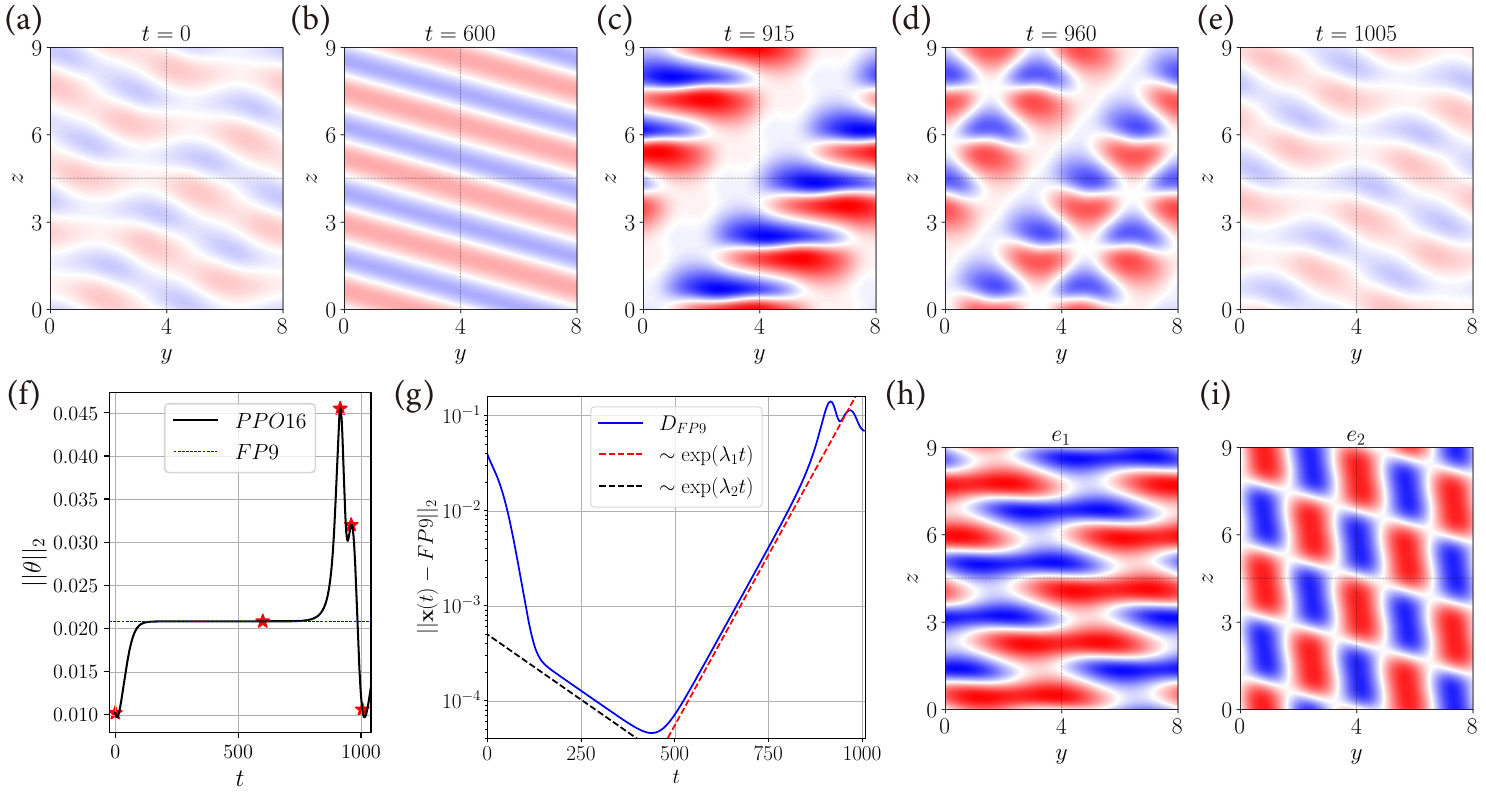}
    \captionsetup{font={footnotesize}}
    \captionsetup{width=13.5cm}
    \captionsetup{format=plain, justification=justified}
    \caption{\label{part3_PO16_series_snapshots} Dynamics of PPO16 with pre-period $T=1007.05$ at $Ra=6240.6429$ (close to the global bifurcation point). (a--e) Snapshots of the midplane temperature field. Snapshot (b) converges to FP9 when used as an initial guess. (f) Time series from DNS. The five red stars indicate the moments at which the snapshots (a)--(e) are taken. (g) $L_2$-distance between each instantaneous flow field of PPO16 and FP9. The dynamics of PPO16 is exponential for most of the cycle (blue curve). The approaching (black dashed line) and escaping (red dashed line) dynamics of PPO16 with respect to FP9 are shown to be governed by two eigenvalues, $\lambda_1$ and $\lambda_2$, of FP9. (h--i) Two eigenmodes $e_1$ and $e_2$ of FP9, visualized via the midplane temperature field.}
\end{figure}

\par Orbit PPO16 is a pre-periodic orbit; its spatial phase shifts by $\braket{\tau(L_y/10, L_z/10)}$ after a pre-period; compare figures \ref{part3_PO16_series_snapshots}(a) and (e). After ten such pre-periods, the final state matches the initial state. The branch of PPO16 states is included in the bifurcation diagram of figure \ref{part3_sepa_BD_PO6_10_14_16_29}. We have continued PPO16 towards increasing $Ra$ to $Ra=6656.54$. Towards decreasing $Ra$, the period of PPO16 increases monotonically and eventually diverges, suggesting a global bifurcation. Figures \ref{part3_PO16_series_snapshots}(a-e) show five snapshots of PPO16 at $Ra=6240.6429$ with pre-period $T=1007.05$, the lowest Rayleigh number we have reached. The corresponding time series in figure \ref{part3_PO16_series_snapshots}(f) indicates that PPO16 slows down significantly between $150\lesssim t\lesssim800$ and spends a long time near an oblique-roll state (figure \ref{part3_PO16_series_snapshots}b). This oblique-roll state is subsequently converged via Newton's method to FP9; figures \ref{part3_PO16_series_snapshots}(b) and \ref{part3_BD-newFP}(g) are related by $\pi_y$. Similarly to PO14 described in \S \ref{part3_UPO14}, PPO16 also bifurcates from FP9 in a global homoclinic bifurcation.

\par We computed the eigenvectors and eigenvalues of FP9 at $Ra=6240.6429$ in the symmetry subspace $\braket{\pi_{y}\pi_{xz}, \tau(L_y/5, L_z/5)}$. The Arnoldi iterations return five leading eigenvalues, all real: $[\lambda_1, \lambda_2, \lambda_3, \lambda_4, \lambda_5] = [0.01651, -0.00631, -0.057, -0.0628, -0.07664]$. Clearly, PPO16 escapes from FP9 along $e_1$, associated with $\lambda_1$ and shown in figure \ref{part3_PO16_series_snapshots}(h), the only unstable eigendirection in this subspace. The direction along which PPO16 approaches FP9 is $e_2$, associated with the second eigenvalue $\lambda_2$ and shown in figure \ref{part3_PO16_series_snapshots}(i). These two eigendirections are subsequently confirmed by subtracting FP9 from PPO16, as well as by the exponential decay and growth rates shown in figure \ref{part3_PO16_series_snapshots}(g). Similarly to the scenario for PO14, here $e_1$ breaks the $O(2)$ symmetry of FP9 in the tilted direction and $e_2$ restores them.

\subsubsection{Orbits RPO21 and RPO22: saddle--node bifurcations}
\label{part3_UPO21_22}
\begin{figure}
    \centering
    \includegraphics[width=\columnwidth]{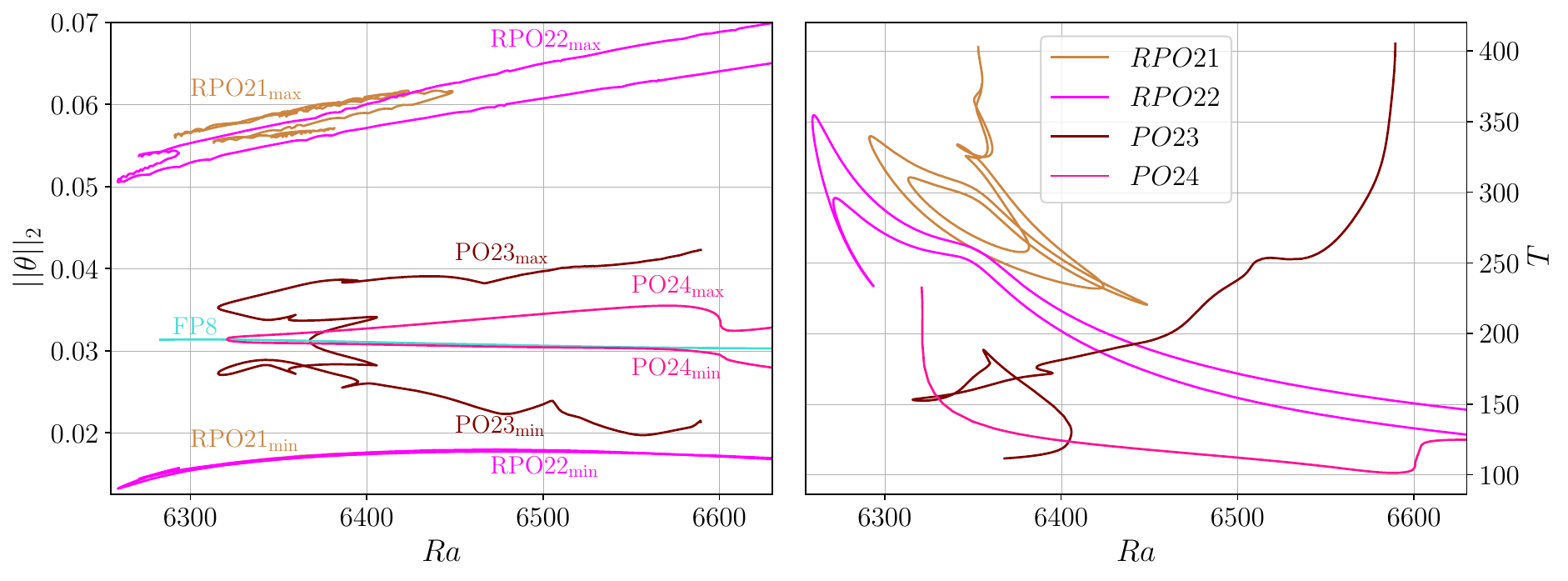}
    \captionsetup{font={footnotesize}}
    \captionsetup{width=13.5cm}
    \captionsetup{format=plain, justification=justified}
    \caption{\label{part3_sepa_BD_PO21_22_23_24} Temperature norms (left) and periods (right) of RPO21, RPO22, PO23 and PO24. On the left, the minima of $\lvert\lvert \theta \lvert\lvert_2$ of RPO21 and RPO22 are too close to be distinguished; the lack of smoothness in the maxima of $\lvert\lvert \theta \lvert\lvert_2$ of RPO21 corresponds to the overtaking of one temporal maximum or minimum of $\lvert\lvert \theta \lvert\lvert_2$ by another as $Ra$ is varied. The creation and termination of RPO21 and RPO22 are not discussed. Both PO23 and PO24 bifurcate from FP8 in two Hopf bifurcations; PO23 possibly terminates in a global bifurcation at $Ra \approx 6589.5$ by meeting FP13, and PO24 exists until at least $Ra=6667$.}
\end{figure}

\par Two relative periodic orbits RPO21 and RPO22 are shown in the bifurcation diagram of figure \ref{part3_sepa_BD_PO21_22_23_24}; both undergo a sequence of saddle--node bifurcations, and the termination and/or creation of both orbits remain unclear. Branch RPO21 exists over the short range $6291.23<Ra<6448.75$; its two endpoints based on our continuation are quite close together, at $(Ra, T) = (6352.99, 402.83)$ and $(Ra, T) = (6354.95, 376.35)$. Integrating RPO21 in time at these two values of $Ra$ does not show remarkable behaviour or a close approach to an equilibrium that would indicate a global bifurcation. Branch RPO22 originates in a saddle--node bifurcation at $Ra=6259$; the upper and lower branches which emanate from it can be continued at least until $Ra=6667$. 

\subsection{Symmetry subspace: reflection}
\label{part3_sym_ref}
\par In this subsection, we discuss two periodic orbits, PO23 and PO24 that are in the symmetry subspace $\braket{\pi_{y}\pi_{xz}} \sim Z_2$. As shown in figure \ref{part3_sepa_BD_PO21_22_23_24}, both bifurcate from FP8 at $Ra=6367.9$ and $Ra=6321$, in two Hopf bifurcations which preserve the reflection symmetry of FP8, and break its translation symmetry $\braket{\tau(L_y/2,0)}$.

\subsubsection{Orbit PO23: Hopf, saddle--node and global bifurcations}
\label{part3_UPO23}
\begin{figure}
    \centering
    \includegraphics[width=\columnwidth]{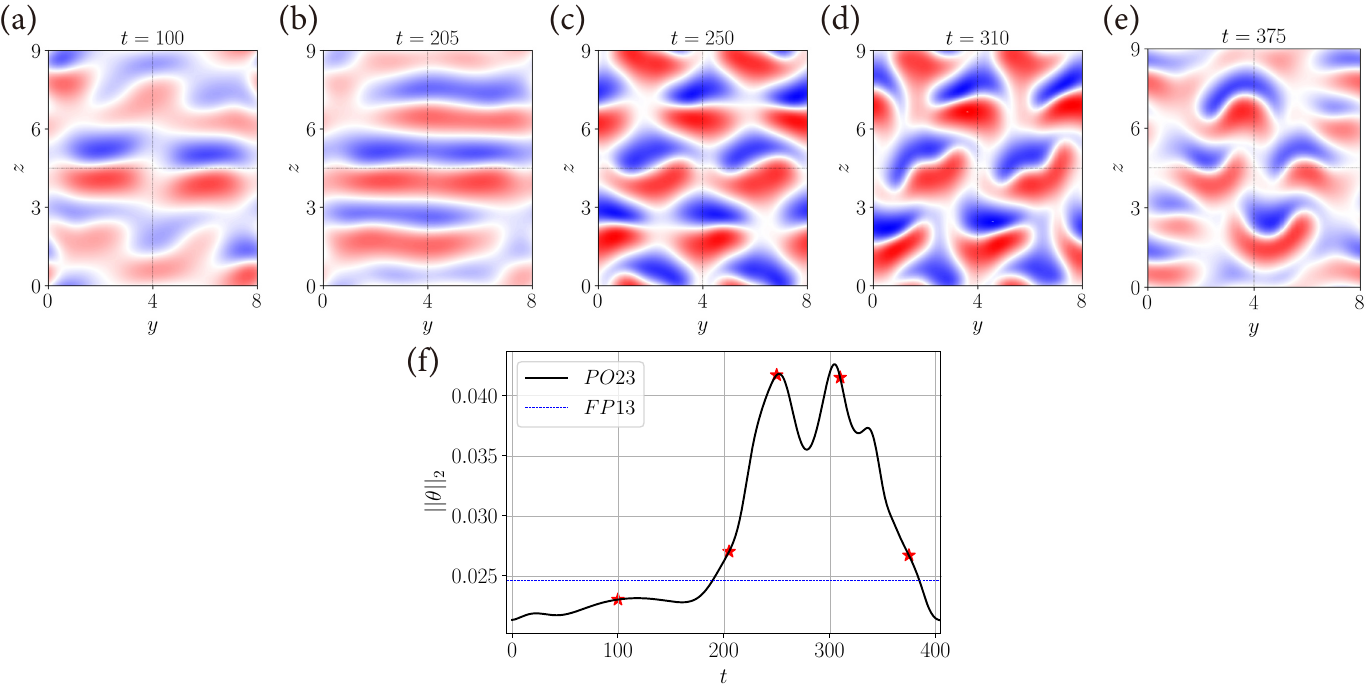}
    \captionsetup{font={footnotesize}}
    \captionsetup{width=13.5cm}
    \captionsetup{format=plain, justification=justified}
    \caption{\label{part3_PO23_series_snapshots} Dynamics of PO23 at $Ra=6589.47$ with period $T=404.6$. (a--e) Snapshots of the midplane temperature field. (f) Time series from DNS. The five red stars indicate the moments at which the snapshots (a)--(e) are taken.}
\end{figure}

\par After its creation, PO23 undergoes a sequence of saddle--node bifurcations. Figures \ref{part3_PO23_series_snapshots}(a-e) show five snapshots of PO23 at $Ra=6589.47$. The initial phase ($25 \lesssim t\lesssim 175$) of PO23 resembles FP13, compare figure \ref{part3_PO23_series_snapshots}(a) and figure \ref{part3_BD-newFP}(k); we used the state in figure \ref{part3_PO23_series_snapshots}(a) as an initial guess for Newton's method to converge to FP13. We have been able to continue PO23 until $Ra=6589.47$, where its period is $T=404.6$. Figure \ref{part3_sepa_BD_PO21_22_23_24} shows that its period seems to diverge; we believe that PO23 terminates on FP13 in a global bifurcation point at a nearby $Ra$.

\subsubsection{Orbit PO24: Hopf bifurcation}
\label{part3_UPO24}
\par After bifurcating from FP8 at $Ra=6321$, PO24 is continued until $Ra=6667$ where we stopped the continuation. It is clear that PO24 oscillates around FP8 and the oscillation amplitude is smaller than that of PO23. We do not show snapshots of PO24.

\subsection{No spatial symmetry}
\label{part3_sym_no_sym}
\par In this subsection, we discuss two orbits without any spatial symmetries.

\subsubsection{Orbit PO9: saddle--node bifurcations}
\label{part3_UPO9}
\begin{figure}
    \centering
    \includegraphics[width=\columnwidth]{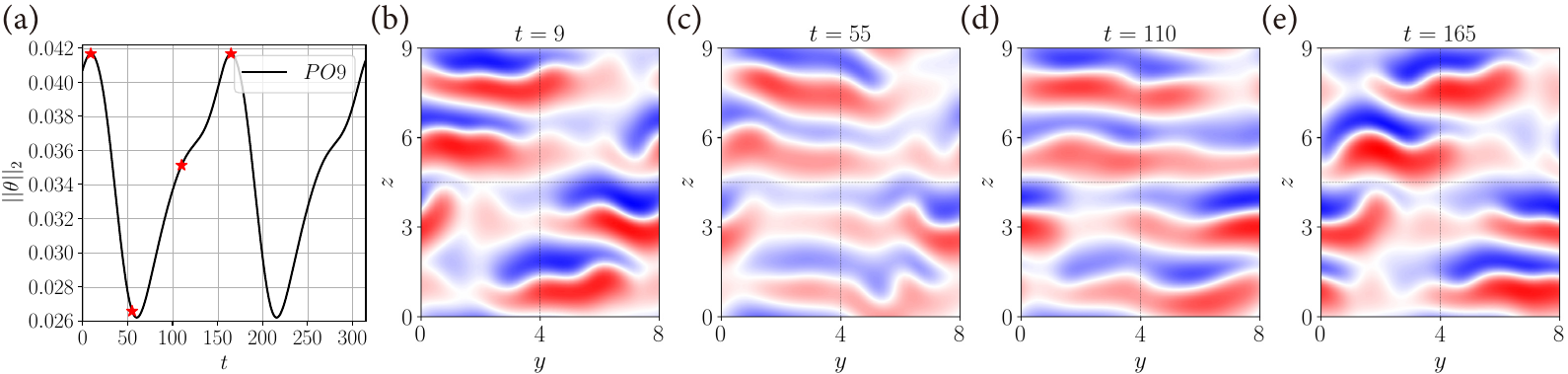}
    \captionsetup{font={footnotesize}}
    \captionsetup{width=13.5cm}
    \captionsetup{format=plain, justification=justified}
    \caption{\label{part3_PO9_series_snapshots} Dynamics of PO9 with period $T=311.18$ at $Ra=6413.11$. (a) Time series from DNS. The four red stars indicate the moments at which the snapshots (b)--(e) are taken.}
\end{figure}

\par Orbit PO9 is shown in the bifurcation diagram of figure \ref{part3_sepa_BD_PO5_7_8_9_30}. Orbit PO9 undergoes a sequence of saddle--node bifurcations. We have continued the lower branch (in period) of PO9 until $(Ra, T) = (6631.4, 188.7)$ and the upper branch until $(Ra, T) = (6475.16, 314.47)$; the origin of PO9 remains unclear. Snapshots and time series of PO9 at $Ra=6413.11$ are shown in figure \ref{part3_PO9_series_snapshots}. Comparing snapshots (b) and (e), we notice that PO9 has the spatio-temporal symmetry
\begin{equation}
(u,v,w,\theta)(x,y,z,t+T/2) = \pi_y\pi_{xz}(u,v,w,\theta)(x, y+0.11L_y, z+0.03L_z, t),
\label{part3_spatiotempo_PO9}
\end{equation}
where $T$ is the period of PO9. Since \eqref{part3_spatiotempo_PO9} contains the combined reflection operator $\pi_y\pi_{xz}$, after two such pre-periods, the discrete translations in $L_{y}$ and $L_{z}$ cancel out and the final state is identical to the initial state---an actual periodic orbit which we converged in our study.

\subsubsection{Orbit RPO11: saddle--node bifurcations}
\label{part3_UPO11}
\begin{figure}
    \centering
    \includegraphics[width=\columnwidth]{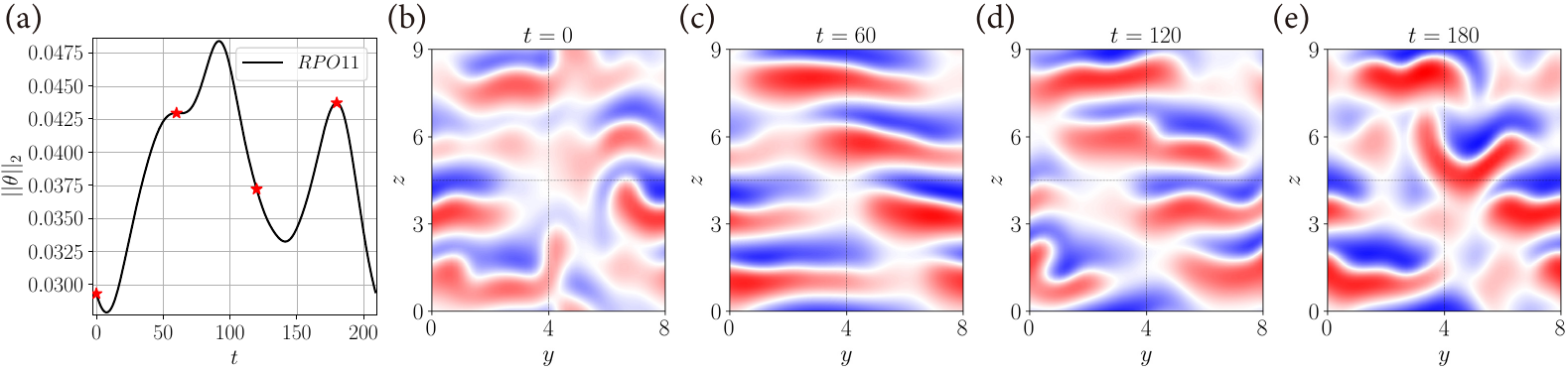}
    \captionsetup{font={footnotesize}}
    \captionsetup{width=13.5cm}
    \captionsetup{format=plain, justification=justified}
    \caption{\label{part3_PO11_series_snapshots} Dynamics of RPO11 with relative period $T=209.26$ at $Ra=6500$. (a) Time series from DNS. The four red stars indicate the moments at which the snapshots (b)--(e) are taken.}
\end{figure}

\par Branch RPO11 is contained in the bifurcation diagram of figure \ref{part3_sepa_BD_PO11_12_20}. Branch RPO11 undergoes a sequence of saddle--node bifurcations. We stopped the continuation at $(Ra, T) = (6680, 212)$ for the lower branch and at $(Ra, T) = (6680, 230.6)$ for the upper branch; its origin remains unknown. Figures \ref{part3_PO11_series_snapshots}(b-e) show four snapshots of RPO11 at $(Ra, T)=(6500, 209.26)$. Like several of the other periodic orbits, RPO11 contains a state (figure \ref{part3_PO11_series_snapshots}c) with four relatively straight but deformed rolls, which breaks along a diagonal fault line (figure \ref{part3_PO11_series_snapshots}d), leading to disordered states (figure \ref{part3_PO11_series_snapshots}e,b), which then reform into approximate rolls (figure \ref{part3_PO11_series_snapshots}c).

\section{Dynamical relevance of unstable periodic orbits and statistical descriptions}
\label{part3_section_PP}
\par In order to illustrate the possible dynamical relevance of the periodic orbits that we have computed and discussed in \S \ref{part3_UPO}, we show in this section some phase space projections and statistical evidence supporting the close visit of orbits by the chaotic dynamics. In this section, we focus on the dynamics and statistics at a fixed Rayleigh number $Ra=6300$ which is about 10\% above the onset of convection. At $Ra=6300$, we identified 19 different periodic-orbit branches. Because of the many saddle--node bifurcations causing branches to zigzag back and forth in $Ra$, several different solutions can be located at the same value of $Ra$ on the same branch. In particular, 34 periodic orbits exist at $Ra=6300$. We use the notation POX$_{n}$ to refer to the $n$-th solution on branch POX.

\subsection{Phase space projections}
\label{part3_phasespaceprojection}
\begin{figure}
    \centering
    \includegraphics[width=\columnwidth]{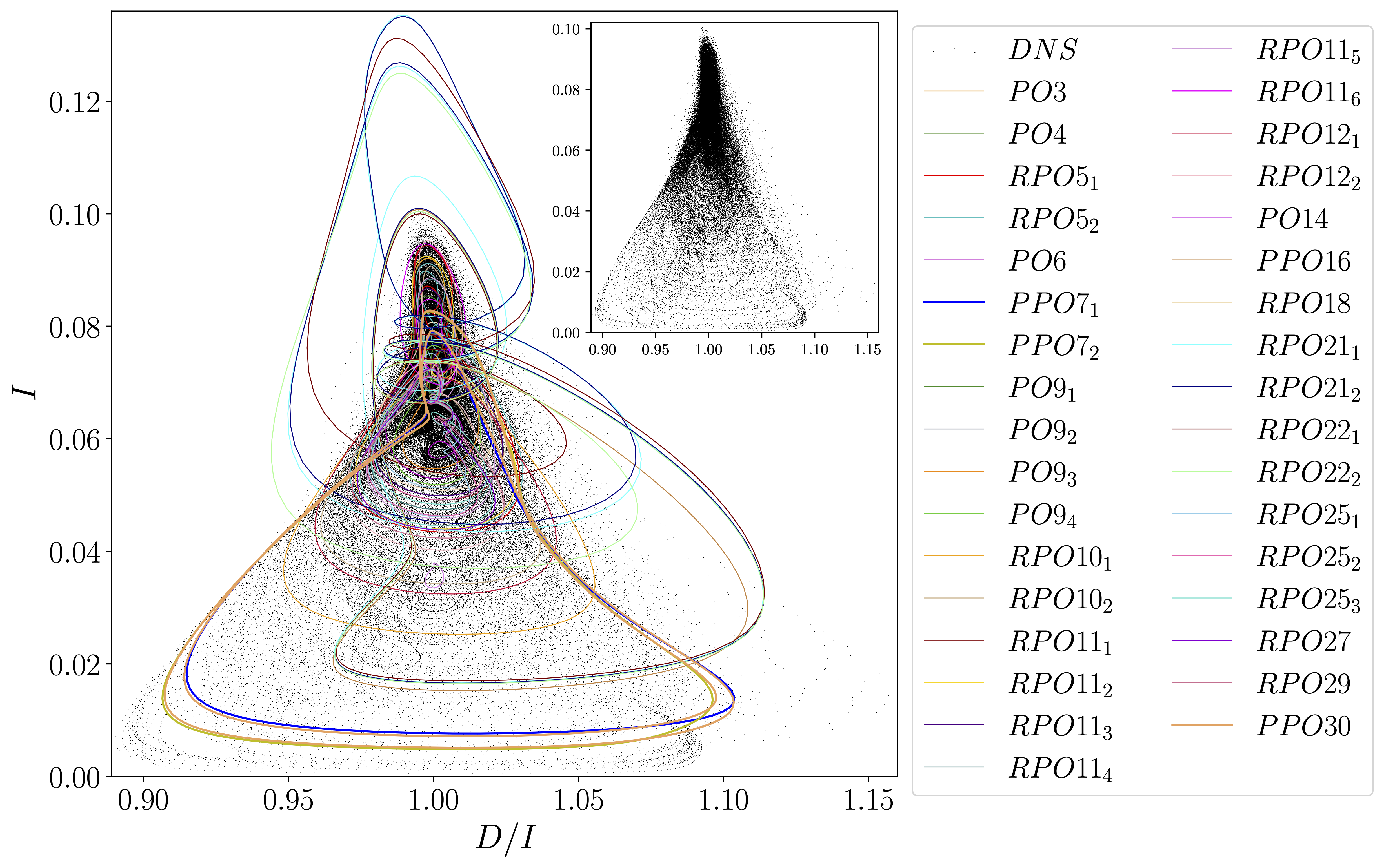}
    \captionsetup{font={footnotesize}}
    \captionsetup{width=13.5cm}
    \captionsetup{format=plain, justification=justified}
    \caption{\label{part3_PP_6300} Phase space projection at $Ra=6300$. The plot shows the projection of instantaneous flow fields, separated by $\Delta t=1$, of the chaotic dynamics (DNS during $2\times 10^5$ time units) and of 34 periodic orbits onto the thermal energy input ($I$) and the viscous dissipation over energy input ($D/I$). The inset shows the chaotic dynamics only. (The dots appear slightly denser due to the inset's smaller size). The subscripts $n$ in POX$_{n}$ indicate different orbits on the same solution branch related by saddle--node bifurcations.}
\end{figure}

\par
\begin{figure}
    \centering
    \includegraphics[width=\columnwidth]{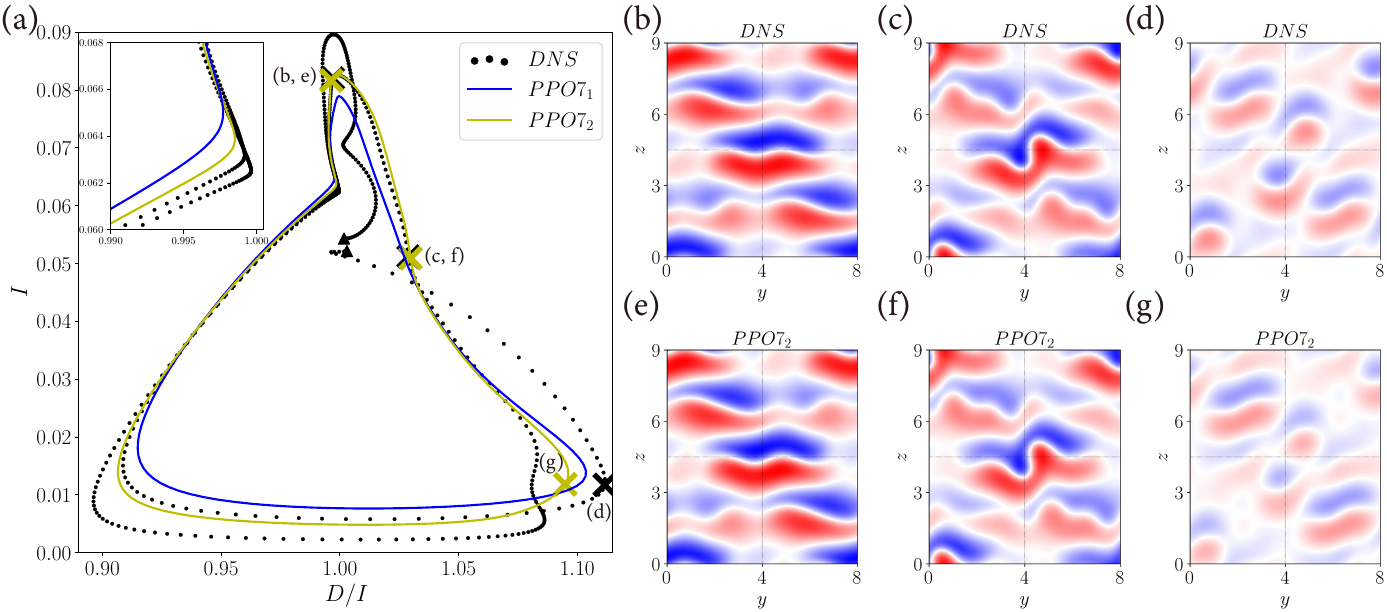}
    \captionsetup{font={footnotesize}}
    \captionsetup{width=13.5cm}
    \captionsetup{format=plain, justification=justified}
    \caption{\label{part3_Projection_PO7} Chaotic dynamics and PPO7 at $Ra=6300$. (a) Projection as in figure \ref{part3_PP_6300} but with only a short portion of DNS and two orbits. The inset zooms in on the slow dynamics close to $D=I$ and $I\approx0.063$. The crosses indicate instants at which the snapshots (b)--(g) are taken and the triangles indicate the beginning and end of the selected DNS trajectory. (b--d) Temperature fields corresponding to three instants of the chaotic dynamics shadowing PPO7$_2$. (e--g) Temperature fields corresponding to three instants of PPO7$_2$.}
\end{figure}

 The projections of a long DNS ($2\times 10^5$ time units, initiated with random small-amplitude noise) and of 34 orbits are shown in figure \ref{part3_PP_6300}. This projection employs two global quantities, the thermal energy input ($I$) and viscous dissipation ($D$), as observables; instantaneous flow fields are represented by dots (for DNS) and closed loops (for orbits) in the $(D/I,I)$ plane. It can be seen that most of the orbits lie on the core part ($0.05\lesssim I \lesssim 0.1$) of the attractor in the current projection. In addition, three orbits shown with thicker curves (PPO7$_{1,2}$ and PPO30) are able to capture some of the very low input events ($I \approx 0.01$) of the DNS trajectory. (Recall in \S \ref{part3_UPO30} that PPO30 bifurcates from the PPO7 branch in period-doubling bifurcations.) To verify that the close match is not merely an artifact of the projection, we examine the flow fields corresponding to these instants. The comparison is shown in figure \ref{part3_Projection_PO7}. In figure \ref{part3_Projection_PO7}(a), we select a short portion of the DNS trajectory (550 time units, belonging to the long DNS trajectory shown in figure \ref{part3_PP_6300}) closely following PPO7$_{1,2}$ in the projection, and we show three temperature fields along the trajectory (marked by crosses in figure \ref{part3_Projection_PO7}a) in figures \ref{part3_Projection_PO7}(b--d). The corresponding flow fields of PPO7$_2$ (which are closest to those of the DNS trajectory both in terms of projection and $L_2$-norm) are shown in figures \ref{part3_Projection_PO7}(e--g). Comparing figures \ref{part3_Projection_PO7}(b,e), (c,f) and (d,g) convincingly suggests that PPO7$_2$ is very closely visited by the chaotic dynamics. (Note that we have shifted the flow fields in $y$ and/or $z$ to facilitate the visual comparison; the optimal shifts are determined by minimizing the $L_2$ difference between two flow fields.)

\begin{figure}
    \centering
    \includegraphics[width=\columnwidth]{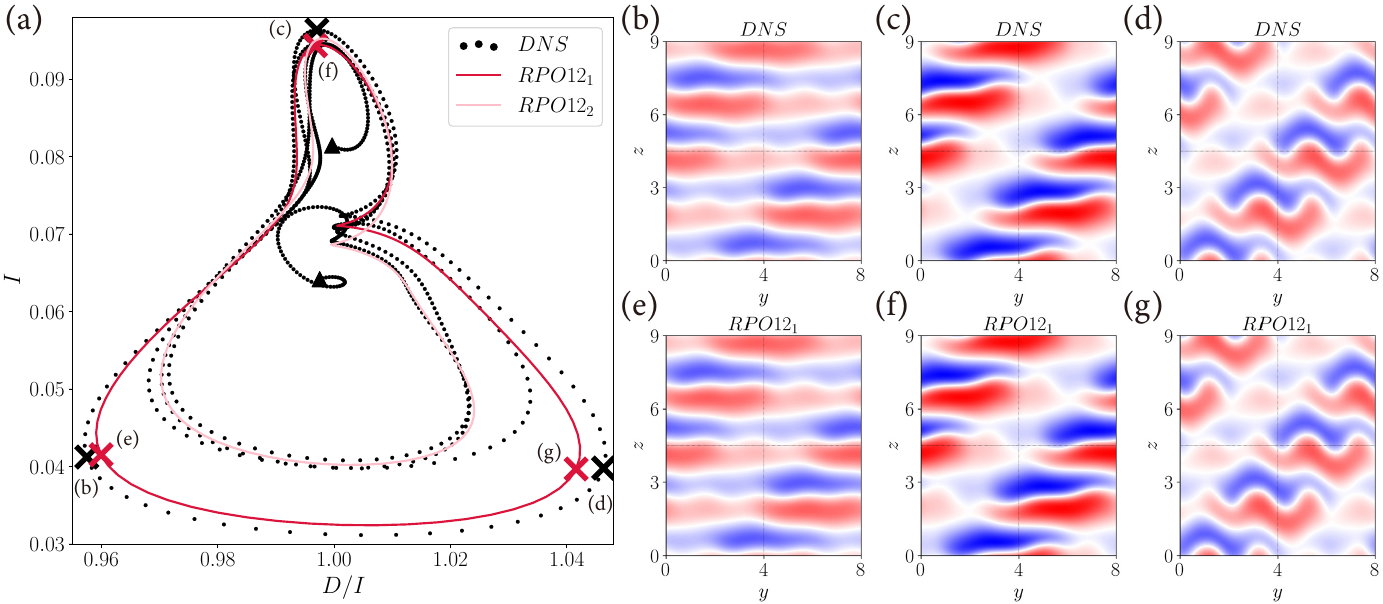}
    \captionsetup{font={footnotesize}}
    \captionsetup{width=13.5cm}
    \captionsetup{format=plain, justification=justified}
    \caption{\label{part3_Projection_PO12} Chaotic dynamics and RPO12 at $Ra=6300$. (a) Projection as in figure \ref{part3_PP_6300} but with only a short portion of DNS  and two orbits. The crosses indicate instants at which the snapshots (b)--(g) are taken and the triangles indicate the beginning and end of the DNS trajectory. (b--d) Temperature fields corresponding to three instants of the chaotic dynamics shadowing RPO12$_1$. (e--g) Temperature fields corresponding to three instants of RPO12$_1$.}
\end{figure}

\par The same analysis can be carried out with almost all other orbits and we will illustrate another example on RPO12 which lies in the core part of the attractor. Figure \ref{part3_Projection_PO12}(a) shows a short DNS trajectory ($830$ time units) approaching RPO12$_1$ and then RPO12$_2$. The almost indistinguishable flow fields between figures \ref{part3_Projection_PO12}(b--d) and (e--g) suggest that RPO12 plays a basic role in the spatio-temporal dynamics of the system. Four orbits (RPO21$_{1,2}$ and RPO22$_{1,2}$) reaching very high input range ($I\approx 0.12$) are not visited at all by the flow, based on the projection in figure \ref{part3_PP_6300}. The reason is that, as stated in \S \ref{part3_sym_ref_5fold}, these four orbits are identified in the symmetry subspace $\braket{\pi_{y}\pi_{xz}, \tau(L_y/5, L_z/5)}$. The five-fold translation symmetry along the domain diagonal greatly constrains the possible dynamics and these four orbits are very unstable in the full phase space.

\subsection{Reconstruction of flow statistics from periodic orbits}
\begin{figure}
    \centering
    \includegraphics[width=\columnwidth]{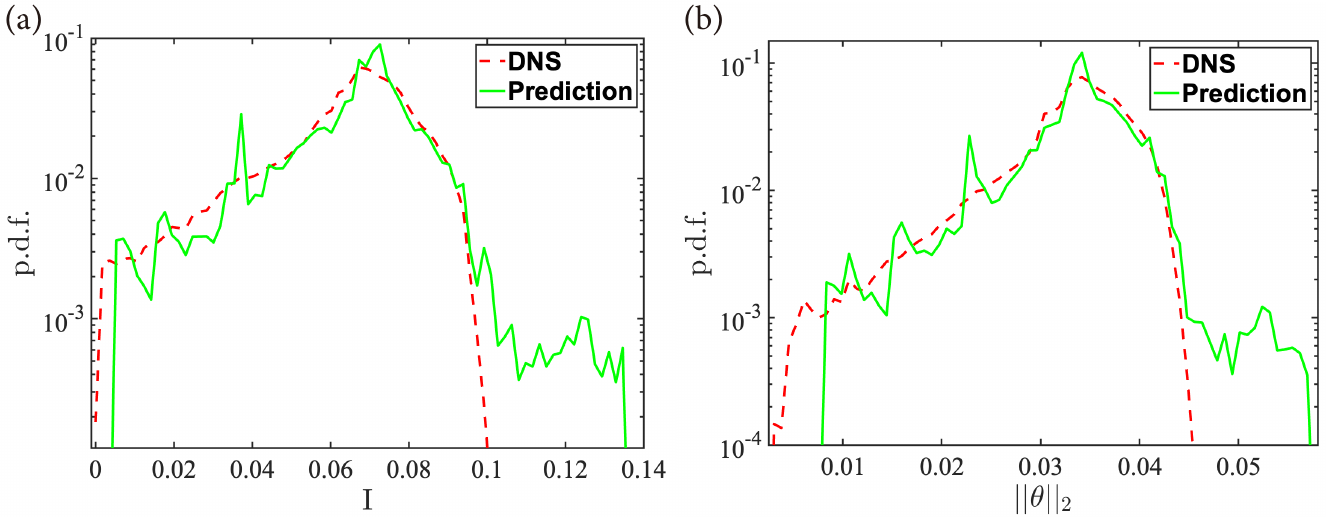}
    \captionsetup{font={footnotesize}}
    \captionsetup{width=13.5cm}
    \captionsetup{format=plain, justification=justified}
    \caption{\label{part3_pdf_I_theta} Probability density functions (PDFs) of $I$ and $\lvert\lvert \theta \lvert\lvert_2$ at $Ra=6300$. Shown are the data from DNS and predicted statistics based on 34 periodic orbits. A total of 80 bins are used for each PDF.}
\end{figure}

\par In \S \ref{part3_phasespaceprojection}, we suggest that many unstable periodic orbits that we have identified are embedded in the trajectories followed by the flow, are closely visited by the chaotic dynamics, and cover the core part of the chaotic attractor. A logical next step is to quantify and to understand how the statistical properties of orbits are related to the statistics of the flow. For this reason, we plot in figure \ref{part3_pdf_I_theta} two probability density functions (PDFs) in terms of the quantities $I$ and $\lvert\lvert \theta \lvert\lvert_2$, reconstructed from 34 orbits at $Ra=6300$, together with the PDF of a long DNS ($2\times10^5$ time units). To reconstruct a PDF from periodic orbits, we use the formula
\begin{equation}
	\Gamma^N_{\text{prediction}} :=\dfrac{\sum_{i=1}^{N}w_i\Gamma_i}{\sum_{i=1}^{N}w_i},
	\label{po_sta_theory}
\end{equation}
where $\Gamma$ is the reconstructed PDF, $N=34$ the number of orbits, $\Gamma_i$ the PDF of the $i^{th}$ orbit, and $w_i$ the weight of the $i^{th}$ orbit. Here, we consider all orbits to be equally important in contributing to the dynamics; in other words, each orbit has the same weight $w_i:= 1$. For other heuristic choices of weights based on the stability and period of orbits, see \S5 of \citet{Chandler2013}.

\par Based on figure \ref{part3_pdf_I_theta}, it can be seen that the peaks of the PDF from DNS, or equivalently the core part of the attractor, are correctly captured by the prediction from periodic orbits. This is consistent with the previous observation on figure \ref{part3_PP_6300}. For both large $I$ and $\lvert\lvert \theta \lvert\lvert_2$ values (rightmost part of each PDF), the predictions show again that some orbits are not visited by the flow, as discussed in \S\ref{part3_phasespaceprojection}. The PDFs in this region are dominated by four (very unstable) orbits (RPO21$_{1,2}$ and RPO22$_{1,2}$), and are 1--2 order(s) of magnitude smaller than the core regions. For $0.005 \lesssim I \lesssim 0.04$ and $0.005\lesssim \lvert\lvert \theta \lvert\lvert_2 \lesssim 0.025$, we observe non-negligible fluctuations in PDFs reconstructed from periodic orbits, while that of the DNS is relatively smooth. Looking back again at the projection in figure \ref{part3_PP_6300}, there are indeed very few identified orbits in the region $0.005 \lesssim I \lesssim 0.04$, suggesting that we are probably missing some important orbits covering these parts of the attractor.

\par Even though the predictions of the statistics (with equal weights) already show reasonably good results, we will leave a deeper study on higher order statistics, different weighting protocols, and eventually periodic orbit theory \citep{Cvitanovic1991} to a future occasion. But as the results in this section suggest, to tackle a quantitative description of transitional turbulence via periodic orbits, a concerted effort is still required to identify a sufficient number of dynamically relevant periodic orbits embedded in the chaotic attractor.

\subsection{Relevance to large domain chaos}
\begin{figure}
    \centering
    \includegraphics[width=\columnwidth]{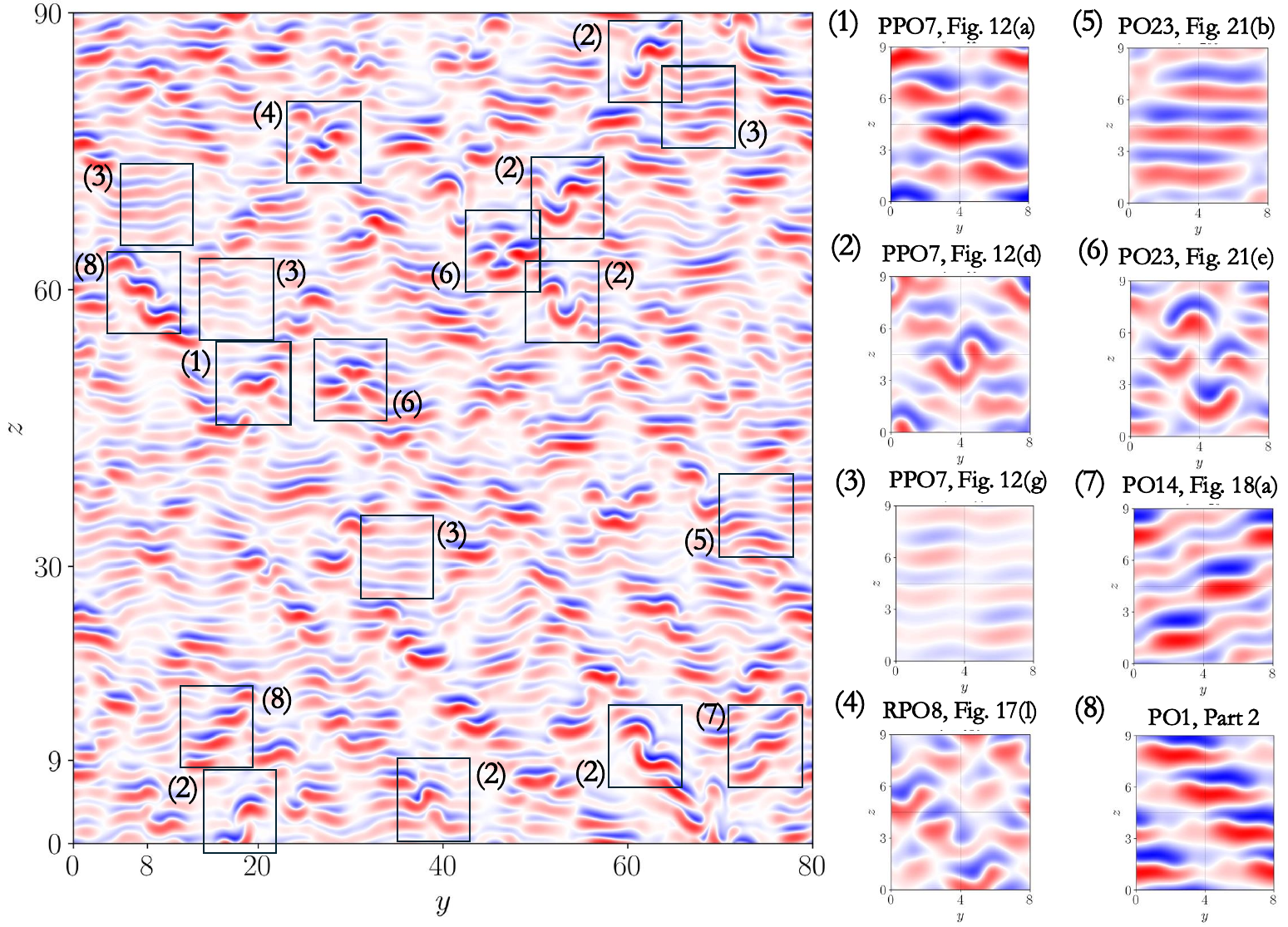}
    \captionsetup{font={footnotesize}}
    \captionsetup{width=13.5cm}
    \captionsetup{format=plain, justification=justified}
    \caption{\label{part3_compa_DNS} Left: one snapshot of temperature field from DNS in a large spatial domain $[L_x, L_y, L_z] = [1, 80, 90]$ at $Ra=6300$ and at a fixed time. Smaller boxes of size $[L_x,L_y, L_z] = [1, 8, 9]$ surround patterns that are (approximately) captured by invariant solutions, shown in the eight snapshots on the right, that have been studied in this work or in \citet{Zheng2024part2}. The grid used for the large domain computation has the same density of points as that for the smaller domain.}
\end{figure}

\par To conclude this section, we show in figure \ref{part3_compa_DNS} (left) an instantaneous temperature field from a chaotic simulation in a large spatial domain of size $[L_x,L_y,L_z]=[1,80,90]$ at $Ra=6300$. Despite the much greater freedom allowed in the large domain, the traces of invariant solutions that we studied in the small domain of size $[L_x,L_y,L_z]=[1,8,9]$ are still present. The small numbered boxes contain portions of the pattern which resemble the equilibria and periodic orbits studied in this work and in \citet{Zheng2024part2}. The eight snapshots in figure \ref{part3_compa_DNS} (right) are labelled by the number of each box and by the corresponding solution names. In particular, the different phases of the roll-bursting dynamics (shown in figure \ref{part3_PO7_series_snapshots} for PPO7) are seen. (Here, we determined these locations by eye, but a more systematic and quantitative approach could also be used.) This correspondence, while qualitative, suggests that unstable solutions in small domains are relevant to the spatio-temporal dynamics in a large domain and provide a promising framework for understanding the bifurcation-theoretic origins of various complex states.

\section{Discussion and conclusions}
\label{part3_conclusion}
\par In this work, we have discussed 26 newly identified periodic-orbit branches in the vertical thermal convection system at fixed Prandtl number $Pr=0.71$ in a fixed-size domain $[L_x,L_y,L_z]=[1,8,9]$ and in the Rayleigh number range $6225\lesssim Ra \lesssim6650$. These new branches, together with four previously studied in \citet{Zheng2024part2}, bring the total to 30 periodic-orbit branches. To the best of our knowledge, this is the largest number of solutions found thus far in three-dimensional Navier--Stokes systems and there certainly exist many more solution branches that we have not followed.

\par The bifurcations that these branches undergo and their spatial symmetries are summarized in table \ref{part3_summary_UPO}; eight different symmetry groups have been identified. Several orbits (RPO8, PO9, RPO11, RPO21 and RPO22) are of unknown bifurcation-theoretic origins in the Rayleigh number range we have studied. Five isolas are found, which might be connected to other branches if other parameters were varied (e.g.\ inclination angle, Prandtl number or domain size); we do not explore this. 

\par As mentioned throughout \S \ref{part3_UPO} and in table \ref{part3_summary_UPO}, almost all of the branches undergo multiple saddle--node bifurcations. Because of this, one branch often contains several portions in the same Rayleigh-number range. Taking this into account, we find that 34 periodic orbits exist at $Ra = 6300$, 45 at $Ra = 6400$, and 23 at $Ra = 6500$, all with different periods, as well as different dynamics and thus different statistics. In addition, there also exist ``ghost states'' that emerge from saddle--node bifurcations \citep{Zheng2025ghost}, which resemble the solutions at the saddle--node bifurcation. Like nearby unstable states, ghosts influence the trajectory of the chaotic dynamics near a saddle--node bifurcation and are relevant for the spatio-temporal patterns observed in weakly turbulent flows.

\par Local (Hopf) and global (heteroclinic or homoclinic) bifurcations are two standard scenarios by which periodic orbits are created or destroyed. We have found three orbits which are born from equilibria in Hopf bifurcations, and eight orbits that disappear or appear via global bifurcations. The orbit period diverges at global bifurcation points, which makes them challenging to compute, but we have been able to continue heteroclinic and homoclinic orbits, as we did in \citet{Zheng2024part2}.

\par In addition to various (complicated) bifurcation scenarios of periodic orbits, in \S \ref{part3_phasespaceprojection} we have presented evidence via phase space projections that suggest close approaches of the chaotic trajectory to certain periodic orbits, highlighting the dynamical relevance of unstable orbits to a chaotic flow. These time-periodic solutions can be said to locally guide the flow trajectory and their statistics. Using the probability density functions of these orbits to reconstruct that of the flow, we find that the core part of the chaotic attractor is well captured by the prediction and that the overall agreement is satisfactory. Finally, we have located versions of our solutions in chaotic simulations in a much larger box. Together, the results in this work emphasize the power of the non-linear dynamical systems approach for shedding light on the convection patterns observed in high-dimensional spatio-temporally chaotic systems, and for quantitatively describing them.

\backsection[Acknowledgements]{We thank O. Ashtari for helpful discussions. We acknowledge E. Knobloch, D. Barkley, and C. Beaume for their valuable comments on the paper. We are grateful to the Scientific IT and Application Support (SCITAS) team of EPFL for providing computational resources and high performance computing (HPC) expertise. We thank the referees for contributing their expertise.}

\backsection[Funding]{This work was supported by the European Research Council (ERC) under the European Union's Horizon 2020 research and innovation programme (grant no. 865677).}

\backsection[Declaration of interests]{The authors report no conflict of interest.}

\appendix
\section{Grid resolution verification}
\label{part3_appendix_grid}
\par In order to verify that our results are independent of the grid resolution, we have reconverged RPO15, RPO17, RPO18, and RPO20---all showing slight lack of smoothness in the curves of bifurcation diagram---by employing the new refined grid ($[N_x, N_y, N_z] = [41,136,136]$) and reconstructed their bifurcation diagrams. The relative differences in $L_2$-norm of temperature field $\lvert\lvert \theta \lvert\lvert_2$, periods and locations (in terms of $Ra$) of saddle--node bifurcations between original and new grids are all below $10^{-4}$.

\bibliographystyle{jfm}

\end{document}